\documentclass[preprint,5p,times,twocolumn]{elsarticle}

\usepackage[utf8]{inputenc}
\usepackage{amsmath}

\usepackage{graphicx}
\usepackage[export]{adjustbox}
\usepackage{subfig}
\usepackage{caption}
\usepackage{booktabs}
\usepackage{threeparttable}
\usepackage{multirow}
\usepackage{tabularx}
\usepackage{balance}

\usepackage{xcolor}
\usepackage{pifont}
\usepackage{fontawesome5}

\biboptions{square,sort&compress,comma,numbers}
\usepackage[breaklinks]{hyperref}
\usepackage{enumitem}
\usepackage[most]{tcolorbox}

\usepackage{xspace}
\usepackage{comment}
\usepackage{url}

\usepackage[acronym,toc,nopostdot,nogroupskip,nomain]{glossaries}
\glsdisablehyper

\newacronym{tc}{TC}{Traffic Classification}
\newacronym{cil}{CIL}{Class Incremental Learning}
\newacronym{fedcil}{FedCIL}{Federated Class Incremental Learning}
\newacronym{sota}{SOTA}{state-of-the-art}
\newacronym{dl}{DL}{Deep Learning}
\newacronym{ai}{AI}{Artificial Intelligence}
\newacronym{win}{WIN}{TCP Window Size}
\newacronym{iat}{IAT}{Inter Arrival Time}
\newacronym{tcpws}{TCP WS}{TCP Window Size}
\newacronym{pl}{PL}{Payload Length}
\newacronym{ps}{PS}{Packet Size}
\newacronym{ttl}{TTL}{Time To Live}
\newacronym{to}{TO}{Traffic Object}
\newacronym{nta}{NTA}{Network Traffic Analysis}
\newacronym{ml}{ML}{Machine Learning}
\newacronym[plural=LLMs, firstplural=Large Language Models (LLMs)]{llm}{LLM}{Large Language Model}
\newacronym{dir}{DIR}{Packet Direction}
\newacronym{sni}{SNI}{Server Name Indication}
\newacronym{tls}{TLS}{Transport Layer Security}
\newacronym{dnn}{DNN}{Deep Neural Network}
\newacronym{sdn}{SDN}{Software Defined Networking}
\newacronym{gpu}{GPU}{Graphics Processing Unit}
\newacronym{genai}{GenAI}{Generative Artificial Intelligence}
\newacronym{rq}{RQ}{Research Question}
\newacronym{adb}{ADB}{Android Debug Bridge}
\newacronym[plural=DNS, firstplural=Domain Name Systems]{dns}{DNS}{Domain Name System}
\newacronym[plural=CDN, firstplural=Content Delivery Networks]{cdn}{CDN}{Content Delivery Network}
\newacronym[plural=CA, firstplural=Certification Authorities]{ca}{CA}{Certification Authority}
\newacronym[plural=1D-CNN, firstplural=1D Convolutional Neural Networks]{onecnn}{1D-CNN}{1D Convolutional Neural Network}

\newif\ifdraft
\draftfalse  %

\usepackage{silence}
\ifdraft
    \newcommand{\annote}[1]{\textcolor{orange}{AN: #1}}
    \newcommand{\rcnote}[1]{\textcolor{teal}{RC: #1}}
    \newcommand{\apnote}[1]{\textcolor{magenta}{AP: #1}}
    \newcommand{\amnote}[1]{\textcolor{olive}{AM: #1}}
    
    \specialcomment{draft}{\begingroup\color{blue}}{\endgroup}

\else
    \newcommand{\annote}[1]{}
    \newcommand{\rcnote}[1]{}
    \newcommand{\apnote}[1]{}
    \newcommand{\amnote}[1]{}
    
    \specialcomment{draft}{\begingroup\color{black}}{\endgroup}

\fi

\specialcomment{rewrite}{\begingroup\color{red}}{\endgroup}

\newcommand{\chatgptt}{$\mathtt{ChatGPT^T}$\xspace}
\newcommand{\chatgptm}{$\mathtt{ChatGPT^M}$\xspace}
\newcommand{\copilott}{$\mathtt{Copilot^T}$\xspace}
\newcommand{\copilotm}{$\mathtt{Copilot^M}$\xspace}
\newcommand{\geminit}{$\mathtt{Gemini^T}$\xspace}
\newcommand{\geminim}{$\mathtt{Gemini^M}$\xspace}
\newcommand{\chatgpt}{$\mathtt{ChatGPT}$\xspace}
\newcommand{\copilot}{$\mathtt{Copilot}$\xspace}
\newcommand{\Gemini}{$\mathtt{Gemini}$\xspace}
\newcommand{\whatsapp}{$\mathtt{WhatsApp}$\xspace}
\newcommand{\telegram}{$\mathtt{Telegram}$\xspace}
\newcommand{\mirage}{$\mathtt{MIRAGE}$\xspace}

\usepackage{wasysym}

\journal{Computer Networks}

\begin{document}

\begin{frontmatter}

\title{From Prompts to Packets: A View from the Network on ChatGPT, Copilot, and Gemini}

\author[unina]{Antonio Montieri}\ead{antonio.montieri@unina.it}
\author[unina]{Alfredo Nascita}\ead{alfredo.nascita@unina.it}
\author[unina]{Antonio Pescap\`e}\ead{pescape@unina.it}

\address[unina]{University of Napoli Federico II, Italy}

\begin{abstract}
\begin{draft}
Generative AI (GenAI) chatbots are now pervasive in digital ecosystems, fundamentally reshaping user interactions over the Internet. 
Their reliance on an always-online, cloud-centric operating model introduces novel traffic dynamics that challenge practical network management. 
Despite the critical need to anticipate these changes in network demand, the traffic characterization of these chatbots remains largely underexplored.
To fill this gap, this study presents an in-depth traffic analysis of \chatgpt, \copilot, and \Gemini used via Android mobile apps. 
Using a dedicated capture architecture, we collect two complementary datasets, combining unconstrained user interactions with a controlled workload of selected prompts for both text and image generation. This dual design allows us to address practical research questions on the distinctiveness of chatbot traffic, its divergence from that of conventional messaging apps, and its novel implications for network usage.
To this end, we provide a multi-granular traffic characterization and model packet-sequence dynamics to uncover the underlying transmission mechanisms.
Our analysis reveals app-/content-specific traffic patterns and distinctive protocol footprints.
We highlight the predominance of TLS, with \Gemini extensively leveraging QUIC, \chatgpt exclusively using TLS~1.3, and characteristic Server Name Indication (SNI) values.
Through occlusion analysis, we quantify the reliance on SNI for traffic visibility, demonstrating that masking this field reduces classification performance by up to $20$ percentage points.
Finally, the comparison with conventional messaging apps confirms that GenAI workloads introduce novel stress factors, such as sustained upstream activity and high-rate bursts, with direct implications for capacity planning and network management.
We publicly release the datasets to support reproducibility and foster extensions to other use cases.
\end{draft}

\end{abstract}


\end{frontmatter}

\newcommand{\AN}[1]{\textcolor{blue}{#1}}
\newcommand{\para}[1]{\noindent \textbf{#1:}}

\section{Introduction}
\label{sec:introduction}

\emph{\gls{genai}} has been one of the most impactful breakthroughs of the last few years, reshaping how people interact through the Internet. 
Advancements in \gls{genai} have led to a growing number of applications leveraging \glspl{llm} for content generation---including text, images, audio, and video---especially in the form of interactive \emph{\gls{genai} chatbots} commonly queried via textual prompts.
The impact of \gls{genai} chatbots has been investigated across a range of domains, such as higher education~\cite{wangsa2024systematic}, teaching and research~\cite{akpan2025conversational,samala2025unveiling}, news article analysis~\cite{dubey2024genai}, and diagnostic processes and healthcare screening~\cite{yim2024preliminary}.
The growing and widespread interest in \gls{genai}-powered solutions extends naturally to the networking domain. 
Key verticals impacted include network design, configuration, and security~\cite{liu2024large,wu2024netllm}, as well as network monitoring and management~\cite{bovenzi2025mapping,liu2025large}.

Such massive adoption is further confirmed by the latest Ericsson Mobility Report (June~2025)~\cite{ericsson2025}, which highlights a marked increase in end-user engagement with \gls{genai} chatbots via mobile apps, significantly impacting mobile network traffic. 
The report also stresses that, in response to the rapid rise in popularity and pervasive use of \gls{genai} apps, both application and communication service providers must anticipate shifts in traffic volume and characteristics. In particular, greater emphasis will need to be placed on uplink capacity and latency, as these factors are expected to become critical performance determinants for \gls{genai}-driven workloads.
According to the AppLogic Global Internet Phenomena Report~\cite{applogic}, these observations are further supported by recent trends, which foresee that \gls{genai} chatbots have the potential to become as ubiquitous as personalized search engines. 
Indeed, only in 2024, more than $7\%$ of fixed-device users and $4\%$ of mobile users actively engaged with these \gls{genai} assistants, and this trend is expected to continue increasing soon.

Among these, the most widely adopted and popular \gls{genai} chatbots~\cite{applogic} include \chatgpt%
\footnote{\url{https://chatgpt.com/}},
\copilot%
\footnote{\url{https://copilot.microsoft.com/}},
and \Gemini%
\footnote{\url{https://gemini.google.com/app}}.
For instance, official Italian Audicom data~\cite{audicom2025} show that in April~$2025$, nearly $9$~million Italians---approximately one-fifth of the country's online population---used \chatgpt (with an average of $7.2$~million monthly users), marking a remarkable $+266\%$ increase compared to April~$2024$. Following, \Gemini and \copilot recorded an average monthly user base of approximately $2.3$~million and $1.9$~million, respectively, with these figures continuing to grow in the first months of $2025$.

Their growing popularity, combined with a cloud-based and always-online operating model, results in substantial network usage, making the characterization of their traffic profiles particularly relevant for network monitoring and management.
Unlike conventional apps, \gls{genai} chatbots can generate workloads that involve sustained high-bandwidth exchanges, bursty transmission patterns, and stringent low-latency requirements.
Furthermore, the reliance on large-scale datacenters for inference introduces additional demands on network infrastructure, with an expected increase in downlink and, especially, uplink requirements towards the end of the decade~\cite{ericsson2025}.
These differences become even more evident when contrasted with the traffic patterns of conventional messaging apps, which typically involve lightweight, user-generated content rather than compute-intensive, model-generated responses.

This paper contributes to this emerging networking scenario by providing a \textbf{comprehensive characterization of network traffic generated by \gls{genai} chatbots} when used via mobile (Android) apps---shortly also referred to as \emph{\gls{genai} apps} in the following.
\begin{draft}
Indeed, despite their popularity and distinctiveness, the literature lacks a detailed characterization of their network footprint. 
While initial studies have begun to explore high-level usage trends~\cite{lyu2024measuring}, a fine-grained analysis of the underlying transport mechanisms and specific traffic patterns of \gls{genai} apps is still missing. 
Bridging this gap is paramount to understanding not only \emph{how much} traffic is generated, but also \emph{how} it is delivered.
In this light, our key point is to investigate the peculiarities of \gls{genai} workloads and understand how they may reshape mobile network usage.
The ultimate goal is to offer valuable insights and practical guidelines for traffic engineering, performance optimization, and network management, addressing the demands of these rapidly growing apps.
\end{draft}

Specifically, we aim to answer the following \textbf{\glspl{rq}}:
\begin{tcolorbox}
[colback=gray!10, colframe=black, title=Research Questions]
\begin{enumerate}[label=\emph{\gls{rq}\arabic*:}]
    \item To what extent do \gls{genai} chatbots exhibit distinctive traffic characteristics across content types, and how do these characteristics manifest when analyzed at different levels of granularity?
    \item How do \gls{genai} chatbots differ in their underlying communication mechanisms, and what patterns emerge from their protocol-level behavior?
    \item How effectively can \gls{genai} chatbots and their generated content be distinguished from one another based on traffic features?
    \item How does the traffic generated by \gls{genai} chatbots compare with that of traditional messaging apps when the same content is exchanged?
\end{enumerate}
\end{tcolorbox}

\begin{draft}
Crucially, answering \gls{rq}1 and \gls{rq}2 regarding traffic characterization and protocol behavior is a necessary prerequisite for understanding the unique nature of \gls{genai} workloads and addressing their implications for traffic classification (\gls{rq}3). 
Furthermore, these analyses lay the groundwork for investigating how such peculiarities differ from the network footprint of traditional messaging apps (\gls{rq}4), providing a comprehensive perspective on the network-level impact of \gls{genai} chatbots.
\end{draft}

\subsection{Paper Contribution}
\label{subsec:contribution}

In line with these objectives, the \textbf{main contributions} of the present work can be summarized as follows:
\begin{draft}
\begin{itemize}

    \item We collect and publicly release $\mathtt{MIRAGE\text{-}GenAI\text{-}2025}$%
    \footnote{\label{foot:dataset}\url{https://traffic.comics.unina.it/mirage/mirage-genai-2025}}, 
    two complementary \gls{genai} traffic datasets using \chatgpt, \copilot, and \Gemini, captured via a dedicated mobile-app traffic collection architecture~\cite{aceto2019mirage}. Leveraging the \emph{generic} dataset generated via unconstrained prompts, we provide a %
    traffic characterization at different granularities across \gls{genai} apps and content types (textual and multimodal responses), analyzing per-trace properties (biflow counts, upstream/downstream packets, traffic volumes, and byte/packet rates) and modeling packet-sequence dynamics via Multimodal Markov Chains to uncover app- and content-specific transmission patterns.

    \item We provide a protocol-level characterization of \gls{genai} app traffic, identifying distinct transport adoption strategies (e.g., \Gemini's use of QUIC vs. \chatgpt's TLS~1.3) and performing a breakdown of TLS versions and specific \gls{sni} fingerprints. Building on this analysis, we evaluate payload-based traffic classification of \gls{genai} apps and generated content, and quantify the contribution of \gls{sni} information for traffic visibility through an occlusion analysis.

    \item Leveraging the \emph{controlled} dataset generated using a fixed prompt set, we compare the transmission of identical content through \gls{genai} chatbots and widely used messaging apps (\whatsapp and \telegram). This comparative analysis uncovers the unique network footprint of \gls{genai} workloads, highlighting notable differences compared to traditional messaging traffic.
\end{itemize}
\end{draft}

The remainder of the manuscript is organized as below. 
Section~\ref{sec:setup} describes the \gls{genai} apps analyzed, generated content, and workloads, and provides details on our traffic collection setup. 
The experimental analysis is presented in Section~\ref{sec:results}. 
Section~\ref{sec:bg} surveys related work. 
Finally, Section~\ref{sec:end} summarizes our findings and outlines directions for future research.

\section{Traffic Collection Setup}
\label{sec:setup}
\begin{figure}
    \centering
    \includegraphics[width=\columnwidth, trim={20 10 20 5}, clip]{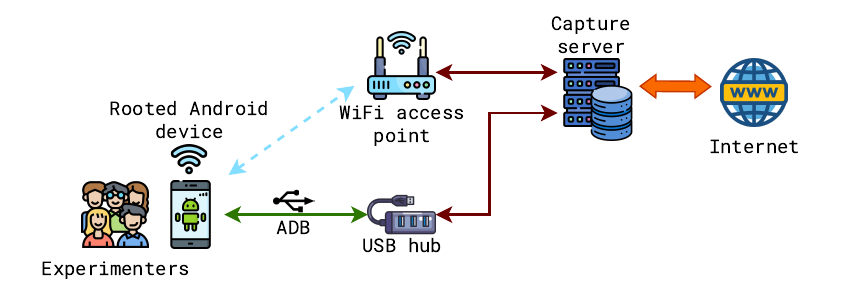}
    \caption{Capture System Architecture. %
    }
    \label{fig:mirage_architecture}
\end{figure}

This section outlines the traffic collection strategy adopted in this work. 
Section~\ref{subsec:apps} describes the \gls{genai} apps and the types of content generated for the \emph{generic} dataset, while Section~\ref{subsec:controlled_prompts} focuses on the dataset obtained through \emph{controlled} prompts. 
Finally, Section~\ref{subsec:capture_setup} details the architecture and setup used for traffic data collection.

\subsection{GenAI Apps and Generic Dataset Generation}
\label{subsec:apps}

\begin{draft}
Given their growing popularity and widespread use~\cite{applogic,audicom2025}, we conduct an in-depth analysis of the network traffic generated by the official Android apps of \chatgpt, \copilot, and \Gemini. 
Regarding the specific subscriptions and models, we employ a \emph{Plus} subscription for \chatgpt (accessing $\mathtt{GPT\text{-}4}$, $\mathtt{GPT\text{-}4o}$, and $\mathtt{GPT\text{-}4o\text{-}mini}$), whereas we use the standard \emph{Free} tier for \copilot (based on a Microsoft customized version of $\mathtt{GPT\text{-}4o}$ and $\mathtt{GPT\text{-}4}$) and \Gemini (based on $\mathtt{Gemini\;1.5\;Flash}$).
Preliminary experiments revealed no statistically significant differences in traffic patterns across model versions within the same app. Consequently, we aggregate the data at the \gls{genai}-app level.
\end{draft}

Notably, these chatbots support a wide range of tasks involving diverse content types, including text, audio, images, and video.
Since different media impose distinct network requirements, we define two specific activity categories: \emph{Textual (\textbf{T})} and \emph{Multimodal (\textbf{M})} content generation. The latter primarily encompasses image generation accompanied by brief textual descriptions.
We selected these two activities as they represent the most frequent interaction patterns with such chatbots.

Notably, these \gls{genai} chatbots are increasingly employed for a wide range of tasks, involving the generation of diverse content types, such as text, audio, images, and video.
Since different media impose varying burdens on a networked environment, we define two kinds of activities: \emph{Textual (\textbf{T})} and \emph{Multimodal (\textbf{M})} content generation, with the second mainly encompassing image generation mixed with brief textual content.
We have chosen these two activities since they reflect the most frequent and common types of interaction with such chatbots.

We underline that in both cases, we use natural language prompts to interact with the chatbots. The difference lies in the type of response requested: in the first case, the prompt is intended to generate a textual reply, whereas in the second case, the prompt explicitly requests an image as part of the response (often accompanied by related descriptive text). 
In this setting, we do not impose any specific control over the prompts used for content generation, aside from specifying whether textual or multimodal output is desired. 
We refer to the resulting workload as the \emph{generic} dataset.

\begin{draft}
All prompts in this dataset are manually generated by \emph{human users} (researchers and students), with no constraints other than the requested generated content type (textual or multimodal).
This approach ensures \emph{ecological validity}, capturing the heterogeneity of real-world human interactions (e.g., varying reading times, thinking pauses, and prompt complexity) similar to established community datasets~\cite{aceto2019mirage, bayat2024itc, zhao2024large}.%
\footnote{\textcolor{black}{We explicitly avoided automated frameworks (e.g., based on ADB) to also prevent triggering the aggressive anti-bot and integrity checks employed by \gls{genai} apps, which could bias the traffic collection.}}
Also, given the long capture duration and the involvement of multiple human experimenters, the resulting traffic is expected to reflect average, representative usage patterns of the considered \gls{genai} apps. 

Notably, this heterogeneity provides the necessary statistical volume and diversity to robustly characterize \gls{genai} apps and contents and to train \gls{ml}-based traffic classifiers (as detailed in Section~\ref{subsec:classification}), ensuring they generalize to legitimate ``wild'' traffic rather than overfitting to specific prompt structures.
On the other hand, for stricter comparative analyses, we rely on the \emph{controlled} dataset described below.
\end{draft}

\subsection{Generation with Controlled Prompts}
\label{subsec:controlled_prompts}

In the \emph{controlled} capture setup, we design a set of \emph{ten different prompts} representing a heterogeneous collection of generative tasks and information retrieval that cover a wide range of thematic domains and cognitive skills. The topics span historical knowledge, technical explanations in computer networking, cultural and gastronomic information, basic mathematical calculations, and simple programming exercises. 
Further tasks involve the search for cultural events in specific geographic locations, the composition of an academic abstract for a paper addressing a specific research problem, translation between languages, and the summarization of complex technical news into concise and accessible statements.
Notably, while the first nine prompts explicitly require a structured textual response in plain text format only, the last prompt requests the creation of a visual illustration, thereby introducing a multimodal generation requirement. 

For this dataset, we use the same \gls{genai} apps considered in the generic dataset, namely \chatgpt, \copilot, and \Gemini. 
Each controlled prompt is executed on all three chatbots to ensure comparability. Unlike the generic dataset, here the prompts are identical across all runs, and the prompt itself determines the content type (textual or multimodal). 
\begin{draft}
All controlled prompts are manually issued by the authors during the capture sessions.
This choice allows us to eliminate prompt-level variability and focus exclusively on application-level behavior.
\end{draft}
To enable a direct comparison with conventional messaging apps, each prompt and generated response (text or image) is subsequently transmitted using \whatsapp and \telegram on the same device and in the same test environment. This approach produces additional traffic traces that correspond to the delivery of the same content via widely used communication apps, allowing for a one-to-one comparison of traffic patterns. 
We refer to the resulting workload as the \emph{controlled} dataset.
The complete list of controlled prompts, along with the corresponding responses from each \gls{genai} chatbot, is publicly available at: \url{https://traffic-arclab.github.io/genai_prompts/}.

\subsection{Traffic Collection Architecture and Setup}
\label{subsec:capture_setup}
The traffic analyzed in this work is collected leveraging the \mirage architecture~\cite{aceto2019mirage} deployed in the ARCLAB laboratory at the University of Napoli Federico II. 
The architecture is depicted in Fig.~\ref{fig:mirage_architecture}. 
In detail, our capture system provides Internet connectivity via a WiFi access point to rooted Android smartphones that generate traffic in response to user inputs, while receiving commands over an off-band USB channel through the \gls{adb}. 
Simultaneous captures via multiple smartphones are managed through their MAC addresses. 
Traffic collection was carried out from October $2024$ to June $2025$ using the latest version of each app available at the beginning of this period. 
We used two rooted Android devices to capture network traffic: a Google Pixel 7a and a Xiaomi Mi 10, both running Android $14$.

\begin{draft}
Each traffic capture results in a PCAP traffic trace and system log files with metadata. 
In detail, to ensure dataset correctness and minimize measurement bias, traffic capture is performed at the WiFi access point (i.e.~first-hop router).
This design prevents potential observer effects introduced by packet capture software running on the smartphone.
However, while router-level PCAP traces provide full visibility of network flows, they do not inherently associate flows with specific apps.
To obtain precise \emph{ground-truth labeling}, we employ rooted smartphones to access system-level networking metadata. 
Specifically, we utilize the \texttt{netstat} tool to periodically map active network sockets to the corresponding process ID. 
This allows us to associate each bidirectional flow (shortly, biflow)%
\footnote{A bidirectional flow or \emph{biflow} is defined as a stream of packets sharing the same 5-tuple (i.e.~transport-level protocol, source and destination IP addresses and ports) regardless of the direction of communication.}
with an Android package name (i.e.~the specific \gls{genai} app), effectively filtering out background traffic generated by the operating system or unrelated apps/services.
Importantly, rooting is used exclusively for this metadata-based labeling and not for encrypted payload inspection.
Also, we deliberately avoid using VPNs or proxy services to simulate different locations, as tunneling protocols introduce encapsulation overhead and alter packet timing, which would bias the fine-grained characterization of original traffic patterns.
\end{draft}

\begin{draft}
As mentioned in Section~\ref{subsec:apps}, the traffic is entirely \emph{human-generated} and was collected with the contribution of researchers and students acting as informed experimenters. 
\end{draft}
During each capture session, the experimenter directly interacted with one of the examined apps. 

For the collection of the \emph{generic} dataset, each capture session lasted $15$~minutes and was dedicated to a single type of content generation, either text or multimodal. 
Overall, the \emph{generic} dataset consists of $20$~hours of traffic for each \gls{genai} app, with $10$~hours dedicated to text generation and the other $10$~hours to multimodal content generation, resulting in a total of $60$~hours of traffic.

\begin{draft}
The \emph{controlled} dataset was obtained by executing the same set of ten prompts on all three \gls{genai} apps, and subsequently transmitting each prompt and the corresponding chatbot response via both \whatsapp and \telegram. 
This additional step ensures that \emph{exactly the same textual content} is exchanged across different apps,
thereby decoupling the impact of the payload from the application-specific network behavior.
As a consequence, any observed traffic differences can be attributed solely to how each app interacts with the network, rather than to variations in the generated content.
All transmissions were performed under identical capture conditions to ensure strict comparability.
\end{draft}
Each controlled capture session lasted $10$~minutes, with one prompt issued per minute, followed by the reception of the complete response before proceeding to the next prompt (after a silence waiting time). 
The first nine prompts requested purely textual output, while the tenth prompt required multimodal content generation. 
Taking into account the additional captures for the \whatsapp and \telegram transmissions, the \emph{controlled} dataset comprises $90$~minutes of traffic in total.

We publicly release the resulting $\mathtt{MIRAGE\text{-}GenAI\text{-}2025}$ datasets to foster replicability and further research on \gls{genai} app traffic.%
\footref{foot:dataset}
Ethical considerations regarding users and user data are briefly discussed in the~\ref{appendix:ethics}.

\section{Experimental Analysis}
\label{sec:results}
\begin{table}[t]
\centering
\footnotesize
\caption{Number of biflows, number of packets, and volume of traffic captured for each GenAI app and content type---text-only (T) or multimodal (M).
For each combination of app and content type, the collected traffic amounts to $10$ hours. 
Packet and volume metrics are reported in terms of absolute values and percentage of downstream traffic.
}
\resizebox{0.95\linewidth}{!}{%
\begin{tabular}{@{}lccccc@{}}
\toprule
& \textbf{Biflows} & \multicolumn{2}{c}{\textbf{Packets}} & \multicolumn{2}{c}{\textbf{Volume}} \\ 
\cmidrule(lr){3-4} \cmidrule(lr){5-6}
& \#  & \# [K] & \% Dwn & [MB] & \% Dwn \\ \midrule
\chatgptt  & $670$  & $1696$  & $69\%$  & $1573$  & $97\%$  \\
\chatgptm  & $716$  & $770$   & $65\%$  & $665$   & $94\%$  \\ \midrule
\copilott  & $911$  & $189$   & $50\%$  & $55$    & $65\%$  \\
\copilotm  & $1004$ & $244$   & $54\%$  & $135$   & $84\%$  \\ \midrule
\geminit   & $532$  & $143$   & $57\%$  & $67$    & $79\%$  \\ 
\geminim   & $229$  & $274$   & $73\%$  & $212$   & $95\%$  \\ 
\bottomrule
\end{tabular}%
}
\label{tab: summary}
\end{table}

\begin{draft}
In the present section, we describe our experimental analysis. 
Before presenting the quantitative results, we provide a high-level overview of the communication process of \gls{genai} apps, which serves as a reference for interpreting the results discussed in the following sections. 
From a network perspective, the interaction with a \gls{genai} app typically consists of a sequence of phases: ($i$) user prompt submission, ($ii$) request transmission to remote backend services, ($iii$) server-side processing and content generation (involving inference latency), and ($iv$) delivery of the generated response to the device. 
Depending on the requested content type, this process may involve a single response exchange (e.g., purely textual) or multiple data transfers associated with multimodal content generation (e.g., images and textual descriptions). Notably, textual responses are often streamed incrementally (i.e.~token-by-token) as they are generated.

This workflow differs from conventional messaging (i.e.~chat) apps, where communication is primarily driven by the exchange of user-generated content and the delivery of complete messages with negligible server-side generation time (i.e.~a store-and-forward model).
The following analysis leverages this conceptual distinction to examine traffic patterns at different granularities and the distinct network footprint of \gls{genai} apps compared to traditional messaging services.

In our setup, each capture session encompasses the entire communication workflow of \gls{genai} apps, from prompt submission to the delivery of the complete response; accordingly, the extracted metrics and features (e.g., protocols and \glspl{sni}) include all traffic exchanges occurring during server-side processing, including intermediate data transfers associated with content generation.%
\footnote{\textcolor{black}{We note that the models considered in this work are not explicit reasoning models. Consequently, they do not exhibit prolonged computation-only or silent phases, and the observed traffic largely corresponds to request transmission and response streaming.}}
Moreover, multiple prompts are sent within a single session to capture successive interaction cycles under consistent network conditions.

The rest of the section is structured as follows. 
\end{draft}
Section~\ref{subsec:characterization} provides an overview of the captured \gls{genai} traffic and a multi-granular characterization at trace and flow levels, including packet-sequence modeling using Multimodal Markov Chains. 
Section~\ref{subsec:protocol} examines the transport and security protocols leveraged by the \gls{genai} apps and presents an in-depth analysis of \gls{tls} biflows. 
Building on this, Section~\ref{subsec:classification} deepens the investigation of \gls{sni} extension adoption through a payload-based traffic classification~\cite{sharma2025survey}, complemented by an occlusion analysis~\cite{nascita2022machine} to quantify the contribution of \gls{sni} information to traffic visibility. 
While these analyses rely on the \emph{generic} dataset, Section~\ref{subsec:controlled} compares upstream and downstream traffic of \gls{genai} chatbots and messaging apps using the \emph{controlled} dataset (i.e.~under identical prompts and responses).
Together, these subsections \emph{directly address the four \glspl{rq}} introduced in Section~\ref{sec:introduction}, each concluding with a concise take-home answer highlighting \emph{actionable insights} for network monitoring, management, and future research.

\begin{figure}[t]
    \centering
    \includegraphics[width=0.5\columnwidth, trim={0 1.5cm 0 1.5cm},clip]{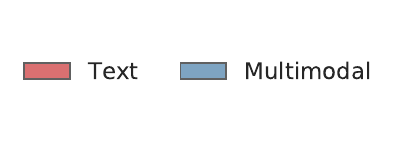}
    \\
    \subfloat[Downstream byte rate {[B/s]}]{%
        \label{fig:boxplots_down_byte}
        \includegraphics[width = 0.48\columnwidth, trim={0 0cm 0 0cm},clip]{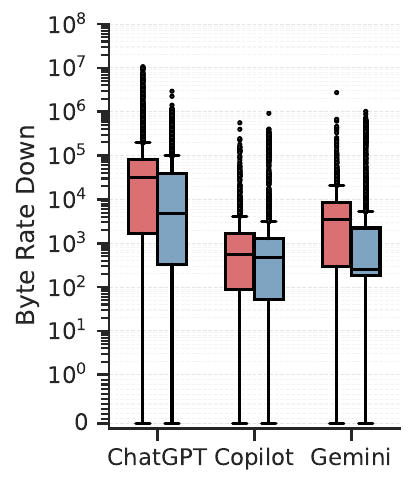}
    }
    \subfloat[Upstream byte rate {[B/s]}]{%
            \label{fig:boxplots_up_byte}

        \includegraphics[width = 0.48\columnwidth, trim={0 0cm 0 0cm},clip]{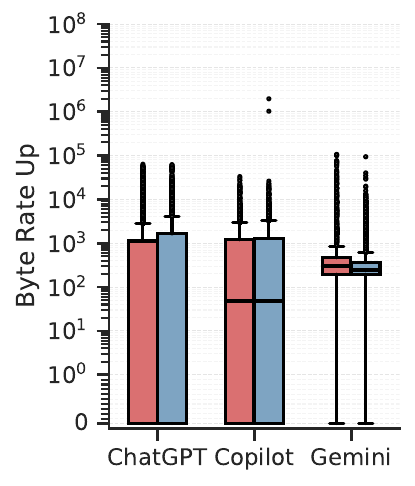}
    }
    \\
    \subfloat[Downstream packet rate {[B/s]}]{%
            \label{fig:boxplots_down_pkts}
        \includegraphics[width = 0.48\columnwidth, trim={0 0cm  0 0cm},clip]{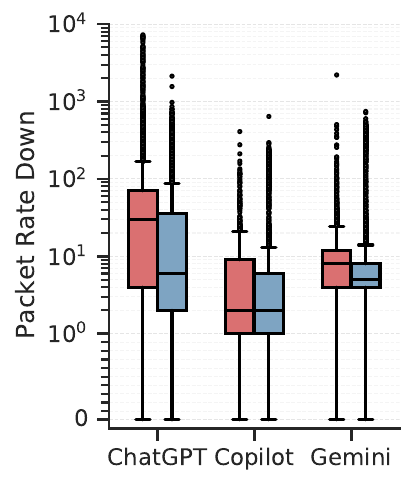}
    }
    \subfloat[Upstream packet rate {[B/s]}]{%
            \label{fig:boxplots_up_pkts}

        \includegraphics[width = 0.48\columnwidth, trim={0 0cm  0 0cm},clip]{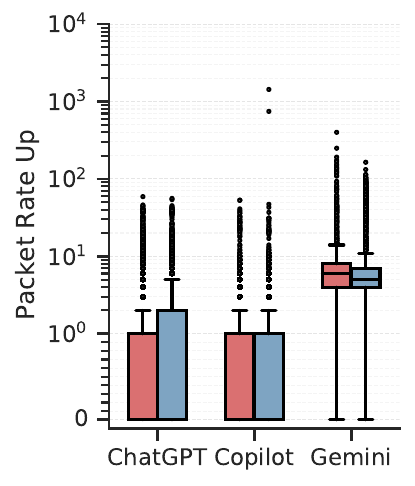}
    }
    \\

    \caption{Downstream byte rate (a), upstream byte rate (b), downstream packet rate (c), and upstream packet rate (d) for different GenAI apps and generated content type. 
    Values are evaluated over time windows of $\Delta = 1\text{s}$ and are displayed in log scale. 
    }
    \label{fig:boxplots}
\end{figure}

\subsection{Multi-Granular Traffic Characterization}
\label{subsec:characterization}

We address \emph{\gls{rq}1} by characterizing \gls{genai} traffic across apps and content types at two granularity levels, trace and flow, highlighting both aggregate behavior and fine-grained dynamics.

\subsubsection{Per-Trace Characterization}
Herein, we provide a quantitative overview of the traffic generated by each \gls{genai} app across the two different response formats (text-only vs.~multimodal). 
Table~\ref{tab: summary} recaps the total number of biflows, total number of packets, and total traffic volume across all capture sessions, broken down by app and content type. 
For completeness, we also report the downstream share for packets and traffic volume. 
We recall that the \emph{generic} dataset analyzed in this section contains $10$~hours of captured traffic for each app and content type.

For \chatgpt and \copilot, the number of biflows is similar regardless of the generated content. Instead, \Gemini produces more than double the biflows during text generation compared to multimodal generation. 
On the other hand, when looking at multimodal content, \Gemini generates the lowest number of biflows, while \copilot accounts for the highest, with more than $1000$ biflows.

In terms of both packets and volume, \copilot and \Gemini generate an amount of traffic for textual responses significantly lower than that generated for multimodal content. 
Detailing, \copilot produces $189$K packets for textual responses and $244$K for multimodal ones ($+29\%$); similarly, the volume increases from $55$~MB to $135$~MB ($+145\%$).
This trend is even more pronounced for \Gemini, with the packet count rising from $143$K to $274$K ($+91\%$) and the traffic volume surging from $67$~MB to $212$~MB ($+216\%$) in the case of multimodal generation.
These results highlight that responses including images cause a significantly higher network load, especially in terms of traffic volume.
\begin{draft}
Conversely, \chatgpt displays a very peculiar behavior. 
Indeed, \chatgptt generates more than twice the number of packets ($1696$K) and traffic volume ($1573$~MB) compared to \chatgptm ($770$K and $665$~MB, respectively). 
A closer examination of \chatgpt's traffic generation---validated through ad-hoc experiments---reveals that this significant inflation stems from the incremental transmission of textual responses (i.e.~streaming) in small batches of tokens (or text chunks). 
While reducing perceived latency, this delivery mechanism fragments the response into a much higher number of small packets. 
Consequently, the cumulative overhead of the protocol headers (e.g., IP, TCP, TLS) attached to each fragment leads to a substantial increase in total traffic volume compared to a standard, non-incremental delivery (e.g., used for images). 
Additionally, the incremental nature of the transmission triggers a higher frequency of control packets (e.g., TCP ACKs) to maintain the stream's flow control and reliability, further inflating the packet count.%
\footnote{\textcolor{black}{This behavior is consistent with the official OpenAI documentation describing the streaming response option (\url{https://platform.openai.com/docs/guides/streaming-responses}). Our tests further confirm that for long textual responses, the protocol-level overhead becomes the primary driver of the observed traffic inflation.}}
\end{draft}

Analyzing the share of downstream/upstream traffic, \copilot shows similar percentages of downstream packets for both types of generated content, while the downstream data volume is higher for multimodal generation than for text generation ($84\%$ vs.~$65\%$). 
For \Gemini, the percentage of downstream traffic is higher than \copilot, both in terms of number of packets and, more significantly, volume, which reaches $95\%$ of total traffic. 
In contrast, the downstream traffic is comparable for \chatgptt and \chatgptm: the number of packets is slightly lower than $70\%$, and the downstream volume exceeds $94\%$ in both cases.

To further deepen the per-trace characterization, we analyze the collected traffic in terms of byte and packet rate. 
Figure~\ref{fig:boxplots} shows the distribution of such transmission rates.
We compute the aggregated rates by considering all traffic sent or received by each app during a capture session, using a window size of $\Delta = 1\ \text{s}$.  
In detail, for a capture starting at $t_0$ with duration $D$, the $i^{th}$ aggregation window considers all the packets sent or received within the interval $[t_0+(i-1)\Delta,t_0 +i\Delta]$, with $i \in \{1,2,\dots,\lceil D/\Delta\rceil \}$. 
Time windows without traffic are excluded from the computation, as they correspond to periods of inactivity by the experimenters.

As illustrated in Fig.~\ref{fig:boxplots_down_byte}, \chatgpt and \Gemini exhibit higher median downstream byte rates when generating textual content compared to multimodal generation. 
In more detail, \chatgpt reaches $\approx\!30$~kB/s for text versus $\approx\!5$~kB/s for multimodal content, while \Gemini reaches $\approx\!3$~kB/s for text and $\approx\!250$~B/s for image generation.
In contrast, \copilot shows comparable median downstream byte rates of $\approx\!500$~B/s for both content types.

Regarding upstream byte rates shown in Fig.~\ref{fig:boxplots_up_byte}, the median value does not change significantly with different generated contents but is significantly different across the apps.
\chatgpt exhibits a median upstream byte rate of $\approx\!0$~B/s, indicating the presence of several time windows containing only downstream traffic, probably due to the effect of incremental transmission mode.
For \copilot and especially \Gemini, this phenomenon is less pronounced (the median upstream byte rate is $\approx\!50$~B/s and $\approx\!250$~B/s, respectively), and likely originates from short-lived responses increasing the number of time windows with both upstream and downstream traffic.

In terms of packet rates (Figs.~\ref{fig:boxplots_down_pkts} and~\ref{fig:boxplots_up_pkts}), the results closely mirror the patterns observed for the downstream/upstream byte rates. Hence, although the absolute values differ, the overall behavior across apps and content types remains consistent.

\begin{tcolorbox}
[colback=gray!10,colframe=black,title=Answer to \gls{rq}1 (per-trace level)]
\gls{genai} chatbots exhibit app- and content-specific differences in network load. 
\copilot and \Gemini show higher load primarily for multimodal generation, whereas \chatgpt exhibits substantial traffic even for textual responses, likely due to the incremental (viz.~token-based) transmission of generated content.
Overall, downstream traffic dominates across all apps; for \chatgpt, it accounts for more than $94\%$ of total volume, regardless of content type.
\end{tcolorbox}

\begin{figure}[t]
    \centering
    \subfloat[{Payload Length [B]}\label{fig:heatmap_pl}]{%
        \includegraphics[width=0.97\columnwidth, trim={0 0 0 0.18cm},clip]{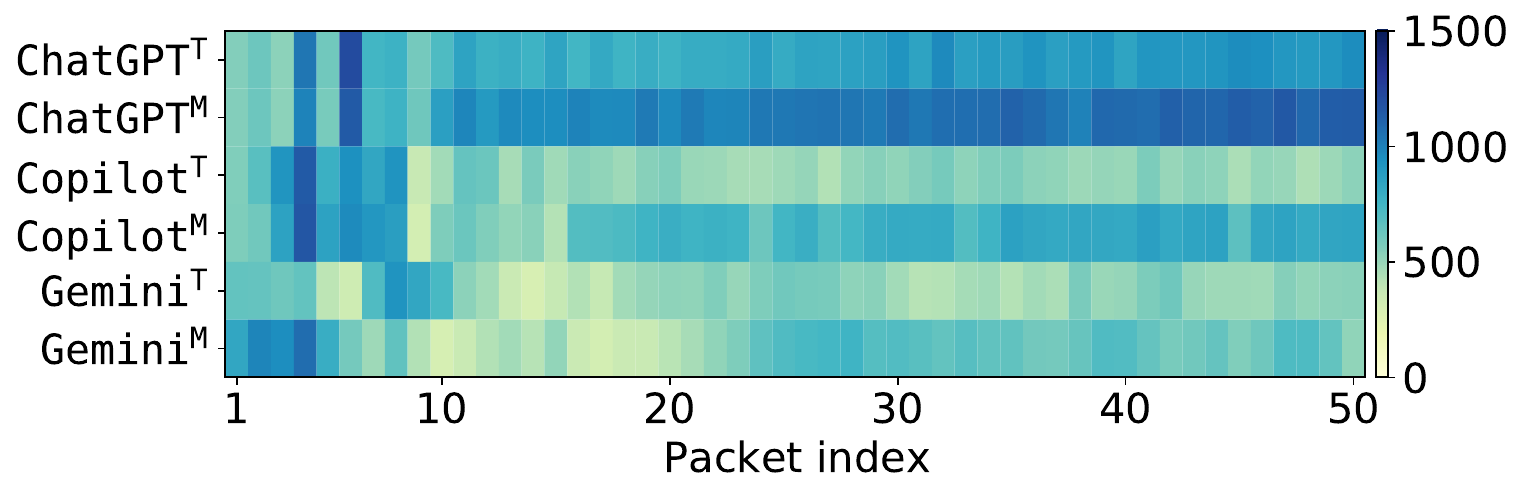}
    }
    
    \subfloat[{Inter-Arrival Time [$\mu$s]}\label{fig:heatmap_iat}]{%
        \includegraphics[width=0.97\columnwidth, trim={0 0 0 0.18cm},clip]{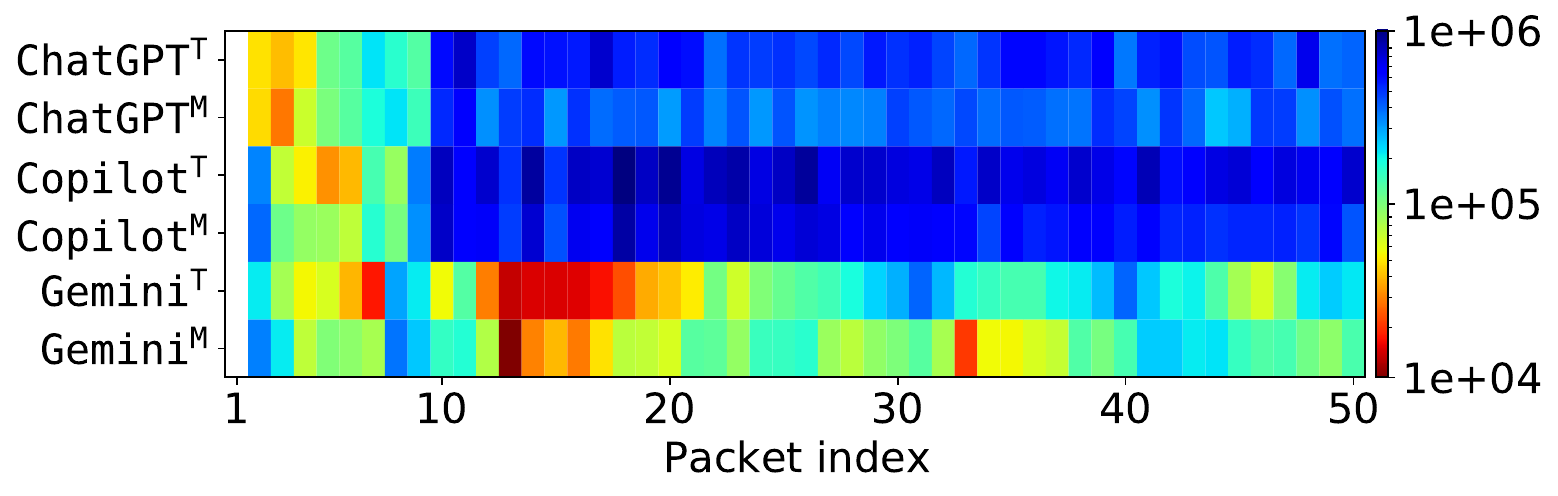}
    }
    
    \subfloat[ Packet Direction\label{fig:heatmap_dir}]{%
        \includegraphics[width=0.97\columnwidth, trim={0 0 0 0.18cm},clip]{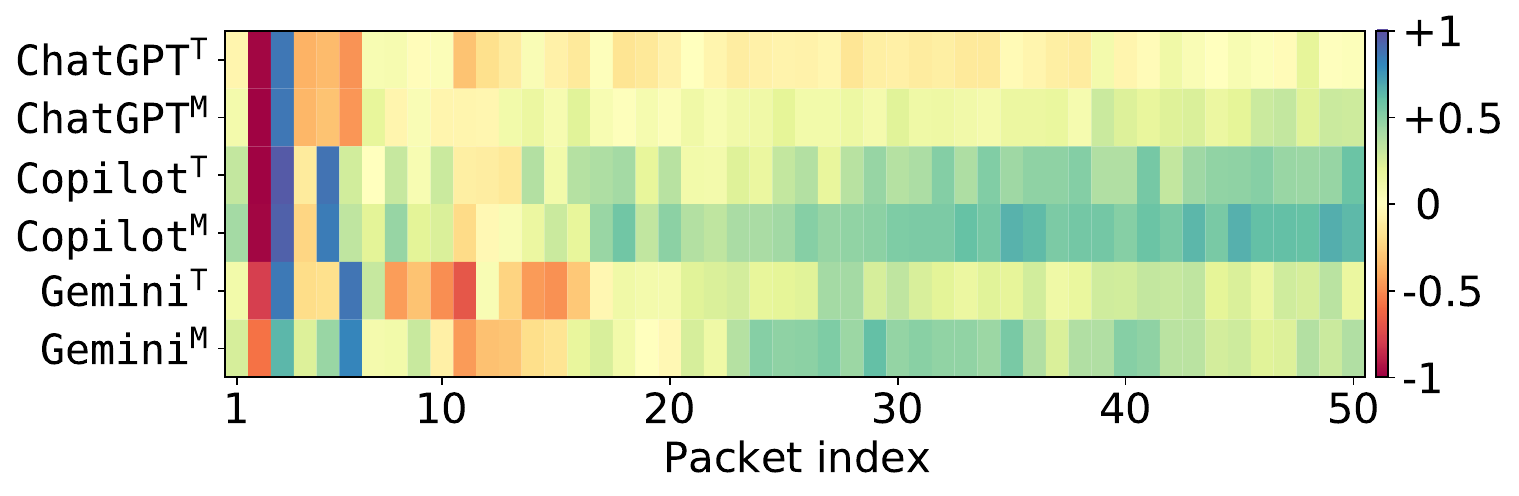}
    }
    \caption{Average of per-biflow PL (a), IAT (b), and DIR (c) time series of the first 50 packets. 
    The IAT is measured in $\mu$s. 
    The downstream and upstream packet direction maps to $+1$ and $-1$, respectively. 
    }
    \label{fig:heatmaps}
\end{figure}

\begin{figure}[t]
    \centering
    \subfloat[\chatgptt]{%
        \includegraphics[width = 0.48\columnwidth, trim={0 0cm 0 0cm},clip]{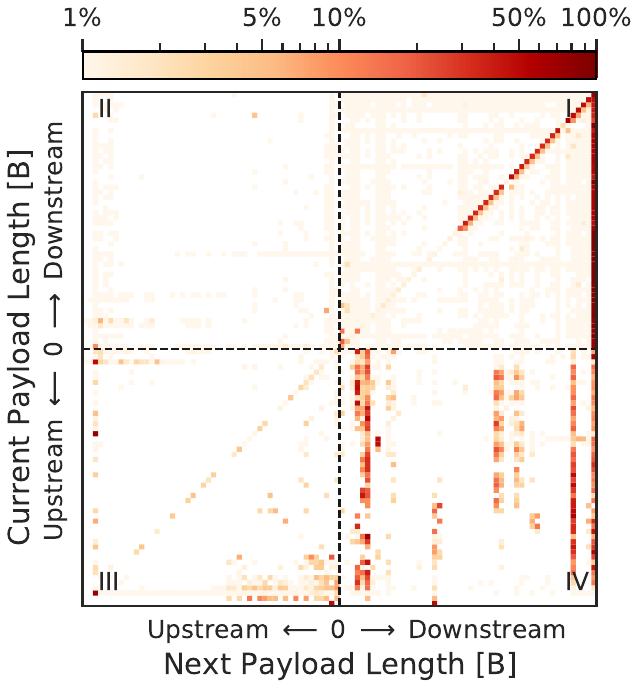}
    }
    \subfloat[\chatgptm]{%
        \includegraphics[width = 0.48\columnwidth, trim={0 0cm 0 0},clip]{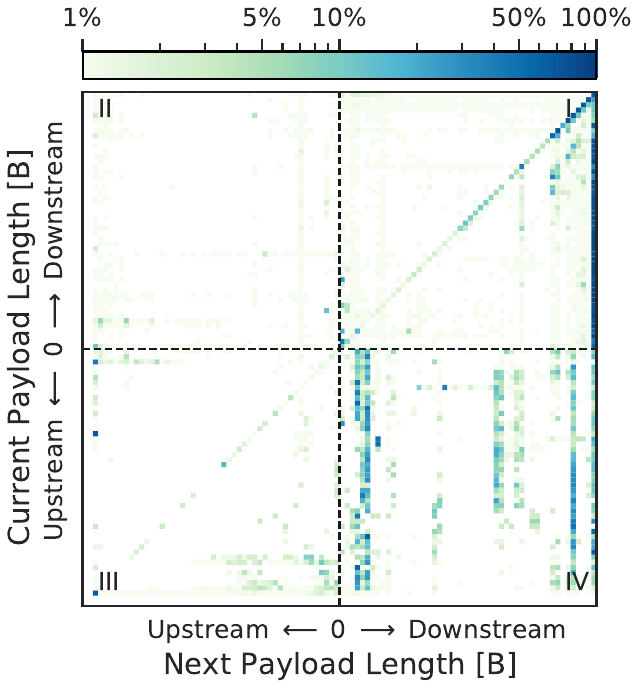}
    }
    \\
    \subfloat[\copilott]{%
        \includegraphics[width = 0.48\columnwidth, trim={0 0cm 0 1.50cm},clip]{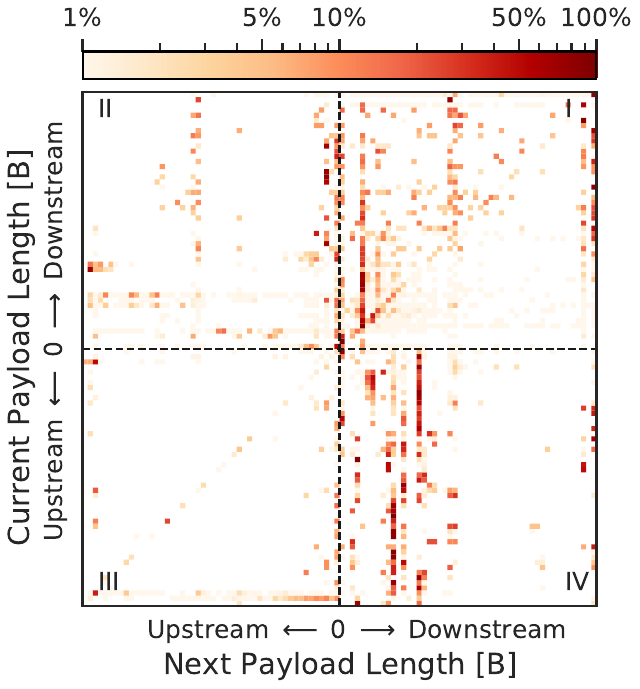}
    }
    \subfloat[\copilotm]{%
        \includegraphics[width = 0.48\columnwidth, trim={0 0cm 0 1.5cm},clip]{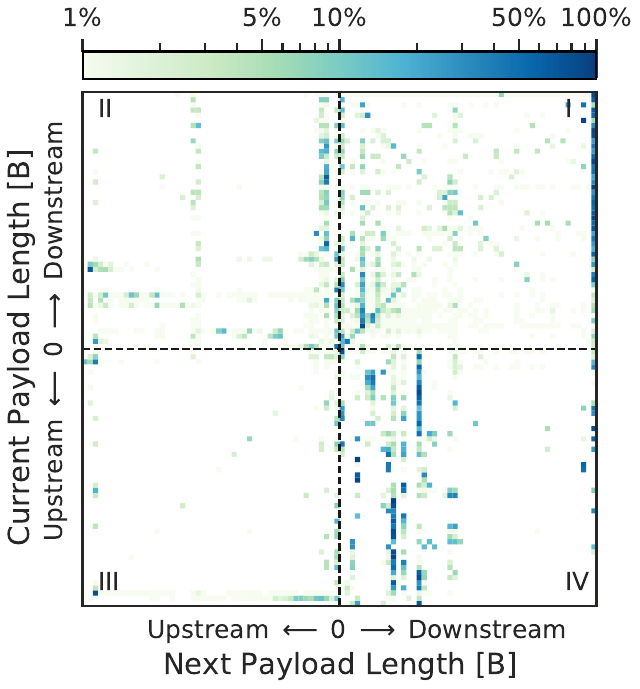}
    }
    \\
    \subfloat[\geminit]{%
        \includegraphics[width = 0.48\columnwidth, trim={0 0 0 1.5cm},clip]{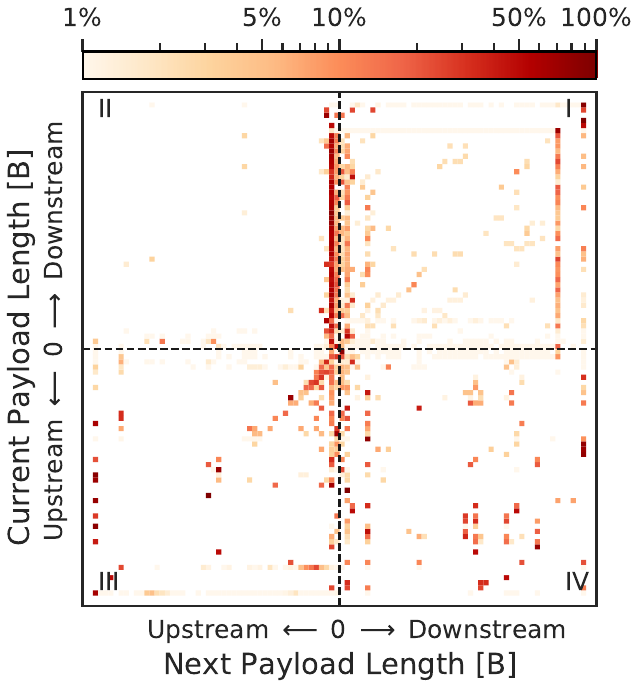}
    }
    \subfloat[\geminim]{%
        \includegraphics[width = 0.48\columnwidth, trim={0 0 0 1.5cm},clip]{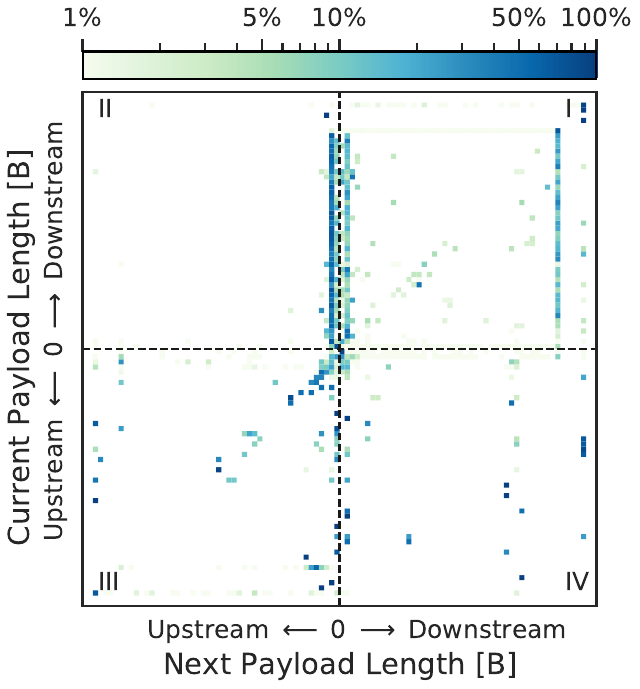}
    }
    \caption{Transition matrices of PL and DIR for \chatgpt (a, b), \copilot (c, d), and \Gemini (e, f). We recall that the matrix quadrants are numbered counterclockwise starting from the top-right.}
    \label{fig:transition_mat}
\end{figure}

\subsubsection{Per-Flow Characterization and Modeling}
Below, we focus on the characterization and modeling of captured traffic at the biflow level, aiming to highlight peculiarities in traffic fingerprint and network behavior across the different \gls{genai} apps and content types. 
Figure~\ref{fig:heatmaps} reports the average values of three time series computed over the first $50$ packets of each biflow. 
Particularly, Fig.~\ref{fig:heatmap_pl} focuses on the \gls{pl}, namely the number of bytes of the transport-level payload; 
Fig.~\ref{fig:heatmap_iat} shows the \gls{iat}, which is the time (measured in $\mu$s) between consecutive packets; 
and Fig.~\ref{fig:heatmap_dir} presents the \gls{dir}, encoded as $+1$ or $-1$ for downstream or upstream packets, respectively. 
From this analysis, we exclude zero-\gls{pl} packets, as they correspond to pure transport-protocol signaling and are not representative of app behavior.

We observe that multimodal content generation exhibits a higher average \gls{pl} compared to text generation. Furthermore, when comparing the different apps, we find that \chatgpt biflows contain larger packets compared to the other two apps.
For the \gls{iat} sequences, no clear trends emerge in the comparison between %
the two types of generated content. 
On the other hand, \Gemini shows shorter \gls{iat} values, particularly for packets after the $10^{th}$.  
Additionally, we observe that the first $9$ packets in \chatgpt and \copilot have lower \glspl{iat} than 
the subsequent ones, whereas \Gemini shows a lower average value around the $15^{th}$ packet, particularly for \geminit. 
When analyzing the \gls{dir}, for all the apps, the first packet is typically upstream and is followed by a downstream response.
Moreover, for multimodal generation, the later packets tend to be more frequent downstream.

To further investigate the characteristics of each app, we model the collected traffic through Multimodal Markov Chains, also distinguishing between different generated contents. 
For each app, we jointly consider the \gls{pl} and \gls{dir} of the packets of all biflows, obtaining a transition matrix in which $<(p_i,d_i),(p_j,d_j)>$ represents the probability that the next packet contains $p_j$ bytes and has direction $d_j$, given that the last observed packet contained $p_i$ bytes and had direction $d_i$. 
To this aim, we first remove all zero-\gls{pl} packets as in the previous analysis, then we discretize the \gls{pl} using an unsupervised binning procedure based on K-means, obtaining $50$ bins as the value balancing quantization error on \gls{pl} versus the number of bins and model complexity.

Figure~\ref{fig:transition_mat} depicts the transition matrices obtained for \chatgpt, \copilot, and \Gemini, considering the generation of text (a, c, e) and multimodal content (b, d, f). 
When comparing the two types of content within the same app, we observe no significant difference for \copilot and \Gemini, while for \chatgptm additional patterns emerge under the main diagonal in the first quadrant, underlining sequences of downstream packets with different but consistently large \glspl{pl}. 

Detailing the behavior of each app, for \chatgptt and (less prominent) \chatgptm only, we spot a vertical line in the first quadrant corresponding to the last \gls{pl} bin, indicating that downstream packets, regardless of their size, are followed by downstream packets with \glspl{pl} close to the maximum size ($1460$~B).
Moreover, the darker upper part of the main diagonal (especially for \chatgptt) suggests that downstream packets with a \gls{pl} larger than $\approx\!650$~B are followed by packets with the same \gls{pl}.
Such patterns disclose the presence of successive downstream packets with different \glspl{pl} as expected from the incremental transmission mode of \chatgptt.
Additional vertical patterns in the fourth quadrant indicate that upstream packets (regardless of their \gls{pl}) are often followed by downstream packets with sizes within certain ranges; particularly, \glspl{pl} close to $\approx\!150$~B, $\approx\!900$~B, $\approx\!1300$~B, or belonging to the last bin (i.e.~with maximum segment size).

On the other hand, \copilot shows more sparse transition matrices, with a higher probability in the upper left triangle of the first quadrant (especially for \copilott), revealing that downstream packets are often followed by other downstream packets of similar or smaller size. 
In addition, it shows vertical patterns in the last bin of the first quadrant, although less pronounced than \chatgpt, and in the left half of the fourth quadrant (i.e.~for \gls{pl} $<\!730$~B).

Notably, \Gemini presents a symmetric vertical pattern in the upper part of the matrix, 
highlighting that downstream packets of different sizes are often followed by small packets (with \glspl{pl} $<\!100$~B) in both directions. 
In addition, the vertical pattern in the right part of the first quadrant indicates that downstream packets are often followed by large downstream packets but with \glspl{pl} smaller than the maximum segment size (i.e.~$\approx\!1250$~B or $\approx\!1380$~B), unlike \chatgpt.

\begin{tcolorbox}
[colback=gray!10,colframe=black,title=Answer to \gls{rq}1 (per-flow level)]
The analysis of per-flow packet sequences shows higher \glspl{pl} for multimodal generation, along with higher \glspl{iat} for \chatgpt and \copilot, particularly after the $10^{th}$ packet, indicating different transmission dynamics. 
Traffic modeling with Multimodal Markov Chains reveals distinctive patterns across \gls{genai} apps, yet similar network behavior across content types within the same app. 
Specifically, \chatgpt primarily generates large \glspl{pl}, especially for successive downstream packets; 
\copilot exhibits a sparser transition matrix, indicating more diverse patterns; 
and \Gemini alternates between either very small or relatively large next \glspl{pl}, regardless of the current \gls{pl}.
\end{tcolorbox}

\begin{figure}
    \centering
    \includegraphics[width=\columnwidth]{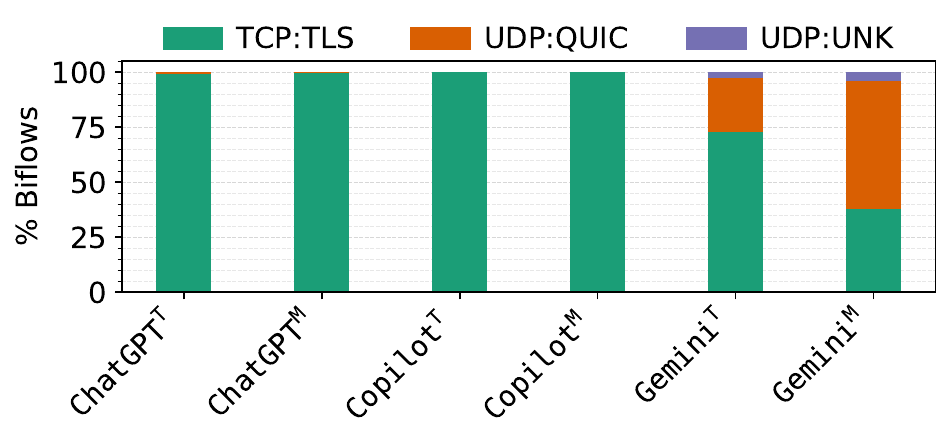}
    \caption{Protocol distribution in terms of biflows of each app and generated content. \textit{UNK} stands for \textit{unknown}.  %
    }
    \label{fig:protocols}
\end{figure}

\begin{table}[htp]
\centering
\caption{Share of biflows ($\mathbf{B}$), packets ($\mathbf{P}$), and volume ($\mathbf{V}$) for each \gls{sni} contacted by \chatgpt, \copilot, and \Gemini during text (T) and multimodal (M) generation. Only \glspl{sni} accounting for more than $1\%$ of the biflows are shown. \glspl{sni} marked with $\ast$ are found in \gls{tls} over QUIC biflows.}
\label{tab:sni}
\resizebox{0.99\linewidth}{!}{%
\begin{tabular}{@{}ccccc@{}}
\toprule
\multirow{2.5}{*}{\textbf{App}} & \multirow{2.5}{*}{\textbf{SNI}} & $\mathbf{B}$ & $\mathbf{P}$ & $\mathbf{V}$\\
\cmidrule{3-5}
& & \multicolumn{3}{c}{[$\%$]}\\
\midrule

\multirow{6}{*}{\rotatebox[origin=c]{90}{\chatgptt}}  & android.chat.openai.com & $76$& $98$& $99$ \\
 & browser-intake-datadoghq.com & $11$& $1$& $<\!1$ \\
 & ab.chatgpt.com & $7$& $1$& $1$ \\
 & chat.openai.com & $3$& $<\!1$& $<\!1$ \\
 & o33249.ingest.sentry.io & $2$& $<\!1$& $<\!1$ \\
 & cdn.jsdelivr.net~$\ast$ & $2$& $<\!1$& $<\!1$ \\
\midrule

\multirow{6}{*}{\rotatebox[origin=c]{90}{\chatgptm}}  & android.chat.openai.com  & $71$& $44$& $36$ \\
 & browser-intake-datadoghq.com  & $9$& $1$& $1$ \\
 & files.oaiusercontent.com  & $8$& $53$& $61$ \\
 & ab.chatgpt.com  & $7$& $2$& $2$ \\
 & chat.openai.com  & $4$& $<\!1$& $<\!1$ \\
 & o33249.ingest.sentry.io  & $1$& $<\!1$& $<\!1$ \\
\midrule

\multirow{7}{*}{\rotatebox[origin=c]{90}{\copilott}}   & gateway-copilot.bingviz.microsoftapp.net  & $35$& $6$& $13$ \\
 & in.appcenter.ms  & $27$& $7$& $14$ \\
 & copilot.microsoft.com  & $14$& $74$& $42$ \\
 & mobile.events.data.microsoft.com  & $9$& $7$& $15$ \\
 & app.adjust.com  & $6$& $1$& $2$ \\
 & graph.microsoft.com  & $4$& $1$& $1$ \\
 & login.microsoftonline.com  & $2$& $<\!1$& $1$ \\
\midrule
 
\multirow{10}{*}{\rotatebox[origin=c]{90}{\copilotm}}  & gateway-copilot.bingviz.microsoftapp.net  & $32$& $4$& $3$ \\
 & in.appcenter.ms  & $25$& $4$& $3$ \\
 & copilot.microsoft.com  & $13$& $29$& $10$ \\
 & mobile.events.data.microsoft.com  & $8$& $5$& $3$ \\
 & app.adjust.com  & $7$& $1$& $<\!1$ \\
 & tse4.mm.bing.net  & $4$& $26$& $36$ \\
 & graph.microsoft.com  & $3$& $1$& $<\!1$ \\
 & tse2.mm.bing.net  & $2$& $13$& $18$ \\
 & tse3.mm.bing.net  & $2$& $12$& $17$ \\
 & tse1.mm.bing.net  & $2$& $6$& $9$ \\
\midrule

\multirow{10}{*}{\rotatebox[origin=c]{90}{\geminit}}  & geller-pa.googleapis.com  & $56$& $6$& $6$ \\
 & proactivebackend-pa.googleapis.com~$\ast$ & $9$& $69$& $58$ \\
 & www.google.com~$\ast$ & $7$& $4$& $8$ \\
 & encrypted-tbn1.gstatic.com  & $5$& $3$& $3$ \\
 & encrypted-tbn0.gstatic.com  & $5$& $2$& $3$ \\
 & encrypted-tbn3.gstatic.com~$\ast$ & $5$& $2$& $2$ \\
 & encrypted-tbn2.gstatic.com~$\ast$ & $3$& $2$& $3$ \\
 & dl.google.com~$\ast$ & $3$& $1$& $1$ \\
 & www.gstatic.com~$\ast$ & $2$& $3$& $5$ \\
 & notifications-pa.googleapis.com~$\ast$ & $1$& $4$& $4$ \\
\midrule

\multirow{11}{*}{\rotatebox[origin=c]{90}{\geminim}}  & lh3.googleusercontent.com~$\ast$ & $27$& $76$& $92$ \\
 & proactivebackend-pa.googleapis.com~$\ast$ & $23$& $21$& $6$ \\
 & geller-pa.googleapis.com  & $10$& $<\!1$& $<\!1$ \\
 & encrypted-tbn3.gstatic.com~$\ast$ & $10$& $<\!1$& $<\!1$ \\
 & encrypted-tbn0.gstatic.com~$\ast$ & $8$& $<\!1$& $<\!1$ \\
 & www.google.com~$\ast$ & $6$& $<\!1$& $<\!1$ \\
 & encrypted-tbn2.gstatic.com  & $6$& $<\!1$& $<\!1$ \\
 & encrypted-tbn1.gstatic.com~$\ast$ & $5$& $<\!1$& $<\!1$ \\
 & assistant-s3-pa.googleapis.com  & $2$& $<\!1$& $<\!1$ \\
 & ssl.gstatic.com~$\ast$ & $1$& $<\!1$& $1$ \\
 & discover-pa.googleapis.com  & $1$& $1$& $<\!1$ \\
\bottomrule

\end{tabular}
}
\end{table}

\subsection{Protocol Stack Characterization}
\label{subsec:protocol}
In this section, we address \emph{\gls{rq}2} by providing a fine-grained protocol-level characterization of the collected traffic across \gls{genai} apps and content types, highlighting distinctive communication patterns. 
We first analyze the composition of collected traffic in terms of the adopted transport and security protocols. 
Then, we investigate specific \gls{tls} extensions to uncover differences across apps that may be useful to discriminate among them. 

Figure~\ref{fig:protocols} depicts the share of biflows for different protocols, categorized by app and generated content. Protocol identification is performed using the \emph{tshark} dissector, and for each biflow, the highest detected protocol in the stack is considered. For clarity, we refer to protocol classes using the TCP/IP stack from the transport layer upward, in the format \texttt{L4:L5}.

We observe that \copilot generates only \gls{tls} biflows, regardless of the generated content, whereas \chatgpt also presents a very limited number of QUIC biflows (likely related to capture sessions including spurious login phase via Google services). 
Differently, \Gemini, in addition to \gls{tls} biflows, also generates QUIC biflows and other UDP biflows indicated with \texttt{UDP:UNK}, for which the dissecting process could not proceed beyond identifying the presence of the transport-layer headers. 
For \Gemini, we also note a difference in the distribution of the protocols depending on the type of generated content. 
In more detail, while the percentage of UDP remains similar,  
\geminim exhibits a substantially higher share of QUIC traffic, reaching $\approx\!60\%$, whereas \geminit accounts for only $25\%$ of QUIC biflows. 
A more in-depth investigation revealed an interesting finding: the identified QUIC traffic actually encapsulates \gls{tls} traffic.%
\footnote{\url{https://datatracker.ietf.org/doc/rfc9001/}}

Since most biflows transport \gls{tls} traffic (whether encapsulated in QUIC or not), we further investigate some \gls{tls} extensions of interest to gain deeper insights.
First, we determine the protocol version using the \emph{ServerHello} packet. 
Our analysis shows that all \chatgpt biflows exclusively use TLS\;1.3, while $\approx\!25\%$ of \copilot biflows still use TLS\;1.2.
Interestingly, for \Gemini traffic, we can distinguish protocol usage based on the type of generated content, suggesting a potential correlation between content type and the chosen \gls{tls} version. Indeed, $73\%$ of \geminim biflows use TLS\;1.3, whereas for \geminit the percentage decreases to $26\%$.

Successively, we analyze the \gls{sni} extension, commonly used as a proxy for traffic identification and classification tasks~\cite{bayat2021deep,akbari2022traffic}.
\gls{sni} is an extension of the \gls{tls} protocol included in the \emph{ClientHello} packet that allows the client to specify which hostname it is attempting to connect to during the handshake phase.
We underline that for \Gemini QUIC biflows, it is still possible to extract this information as the first \emph{Initial} packet has a \emph{CRYPTO} frame containing the \gls{tls} \emph{ClientHello} packet. 

Table~\ref{tab:sni} summarizes the shares of traffic related to each \gls{sni} extracted, in terms of biflows, packets, and volume. 
We observe that there are no common \glspl{sni} between the apps (except for \emph{ingest.sentry.io}) among those accounting for more than $1\%$ of the biflows. 
This suggests that the apps do not share services, thus making the \gls{sni} values potentially useful for recognizing the traffic of different \gls{genai} apps. 

For \chatgpt, \emph{android.chat.openai.com} is the most frequently occurring \gls{sni}, appearing in $76\%$ and $71\%$ of the biflows generated by \chatgptt and \chatgptm, respectively. 
Notably, this \gls{sni} contributes to $\approx\!98\%$ of the total volume for \chatgptt, whereas it represents only $36\%$ of the volume for multimodal generation.
For the latter, the largest volume share (i.e.~$61\%$) is associated with the \gls{sni} value \emph{files.oaiusercontent.com}, which can be exclusively linked to image generation, as it is not present in the traffic generated by \chatgptt.
Additionally, regardless of the generated content, $\approx\!10\%$ of the biflows contact \emph{browser-intake-datadoghq.com}, which provides analytics and performance measurements.

As for \copilot, \mbox{\emph{gateway-copilot.bingviz.microsoftapp.net}} is the most contacted \gls{sni} for both \copilott and \copilotm, accounting for $35\%$ and $32\%$ of biflows, respectively.
Interestingly, the pool of \glspl{sni} \emph{tse*.mm.bing.net} is exclusively associated with image generation, and cumulatively their packet count and traffic volume account for $57\%$ and $80\%$, respectively.

For \geminit, $56\%$ of the biflows contact \emph{geller-pa.googleapis.com}, but \emph{proactivebackend-pa.googleapis.com} represents $58\%$ of its traffic volume with only $9\%$ of the biflows. 
Similarly to \chatgptm, also \geminim uses an \gls{sni} found only for image generation (\emph{lh3.googleusercontent.co}), which accounts for $27\%$ of the biflows, $76\%$ of the packets, and $92\%$ of the traffic volume.

\begin{tcolorbox}
[colback=gray!10,colframe=black,title=Answer to \gls{rq}2]
Our analysis reveals significant differences in the protocol mix and \gls{sni} values across \gls{genai} apps and content types.
\chatgpt and \copilot rely almost exclusively on \gls{tls}, while \Gemini also leverages QUIC. 
\gls{tls} versions likewise diverge: \chatgpt completely moved to \gls{tls}\;1.3, whereas $\approx\!25\%$ of \copilot biflows still negotiate \gls{tls}\;1.2, and \Gemini's version choice varies with content type. 
\gls{sni} values are app- and content-specific with virtually no high-share overlap, and could therefore serve as useful identifiers for distinguishing apps and generated content.
\end{tcolorbox}

\begin{figure}[t]
    \centering
    
    \subfloat[\centering App Classif.\label{fig:cm_app}]{%
    \includegraphics[height = 0.2175\textwidth, trim={0cm 0cm 4cm 0cm}, clip]{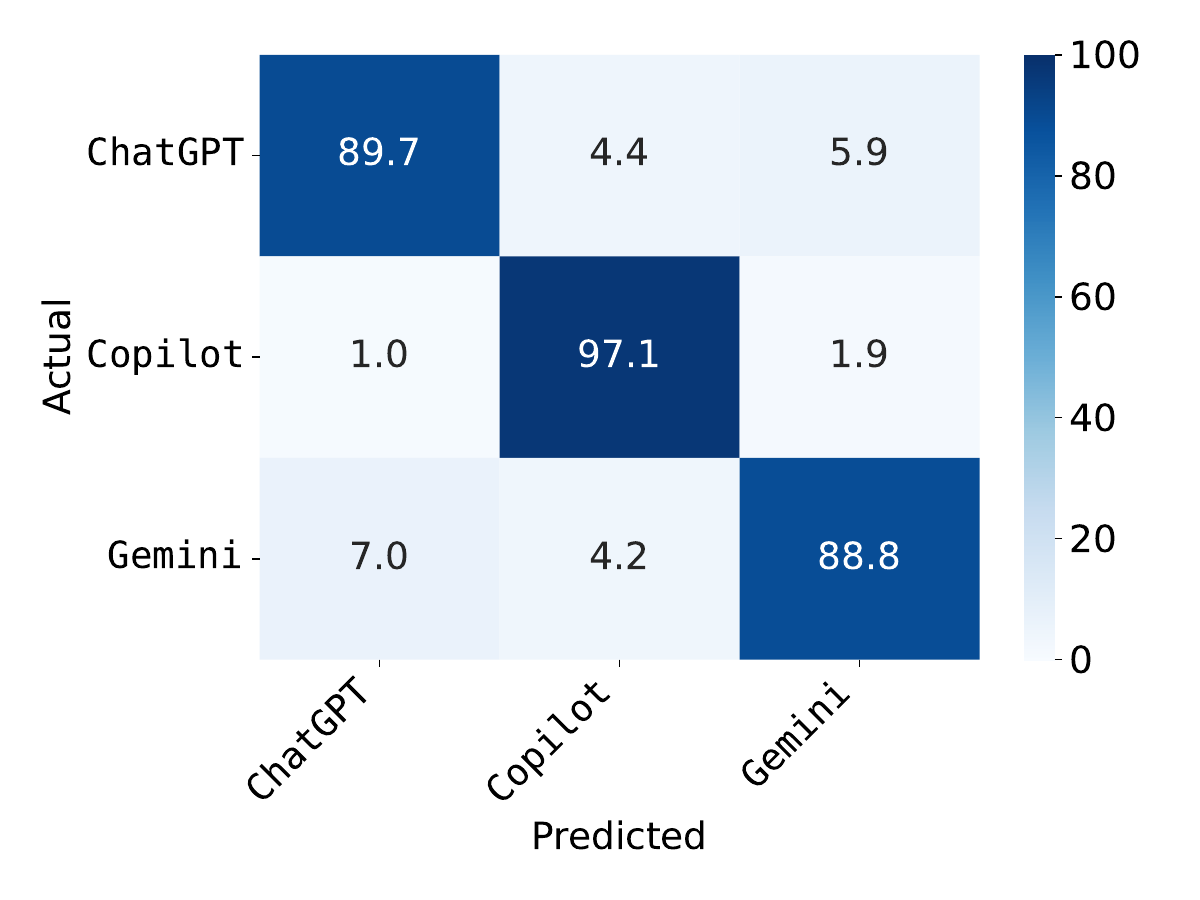}
    }
    \subfloat[\centering App Classif. \\ (masked SNI)\label{fig:cm_app_masked}]{%
        \includegraphics[height = 0.2175\textwidth, trim={2.5cm 0cm 1cm 0}, clip]{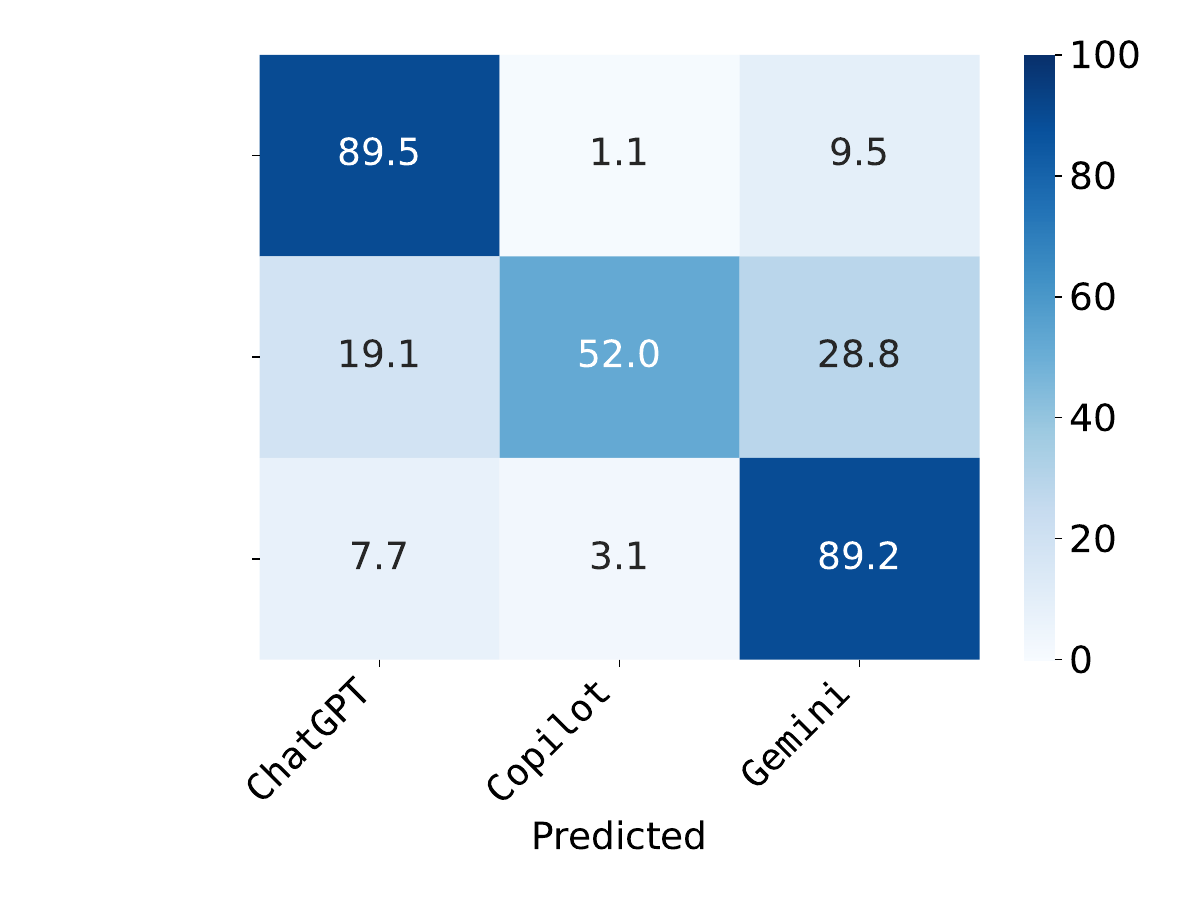}
    }
    
    \subfloat[\centering App\&Activity Classif.\label{fig:cm_app_act}]{%
            \includegraphics[height = 0.22\textwidth, trim={0 0cm 3.5cm 0cm}, clip]{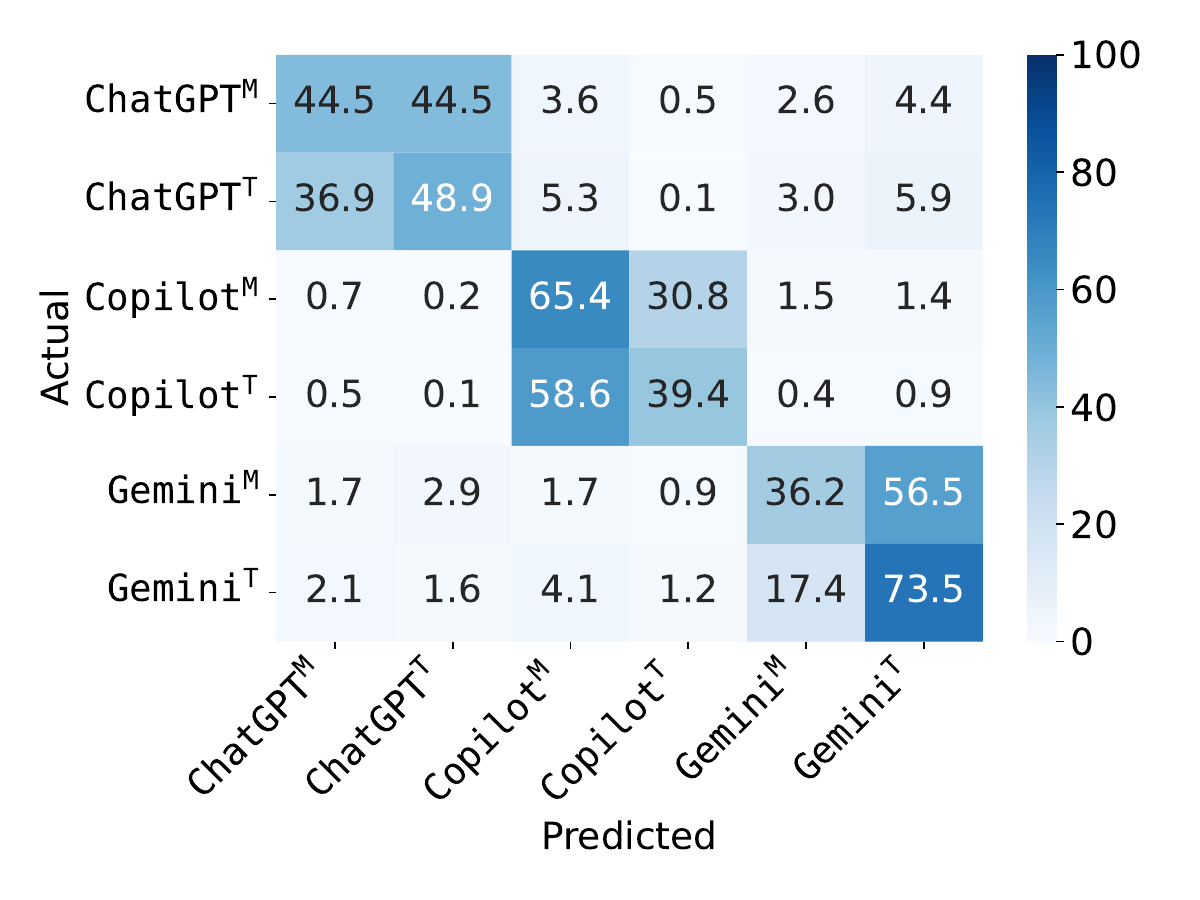}
    }
    \subfloat[\centering App\&Activity Classif. \\ (masked SNI)\label{fig:cm_app_act_masked}]{%
        \includegraphics[height = 0.22\textwidth, trim={3.2cm 0cm 1cm 0}, clip]{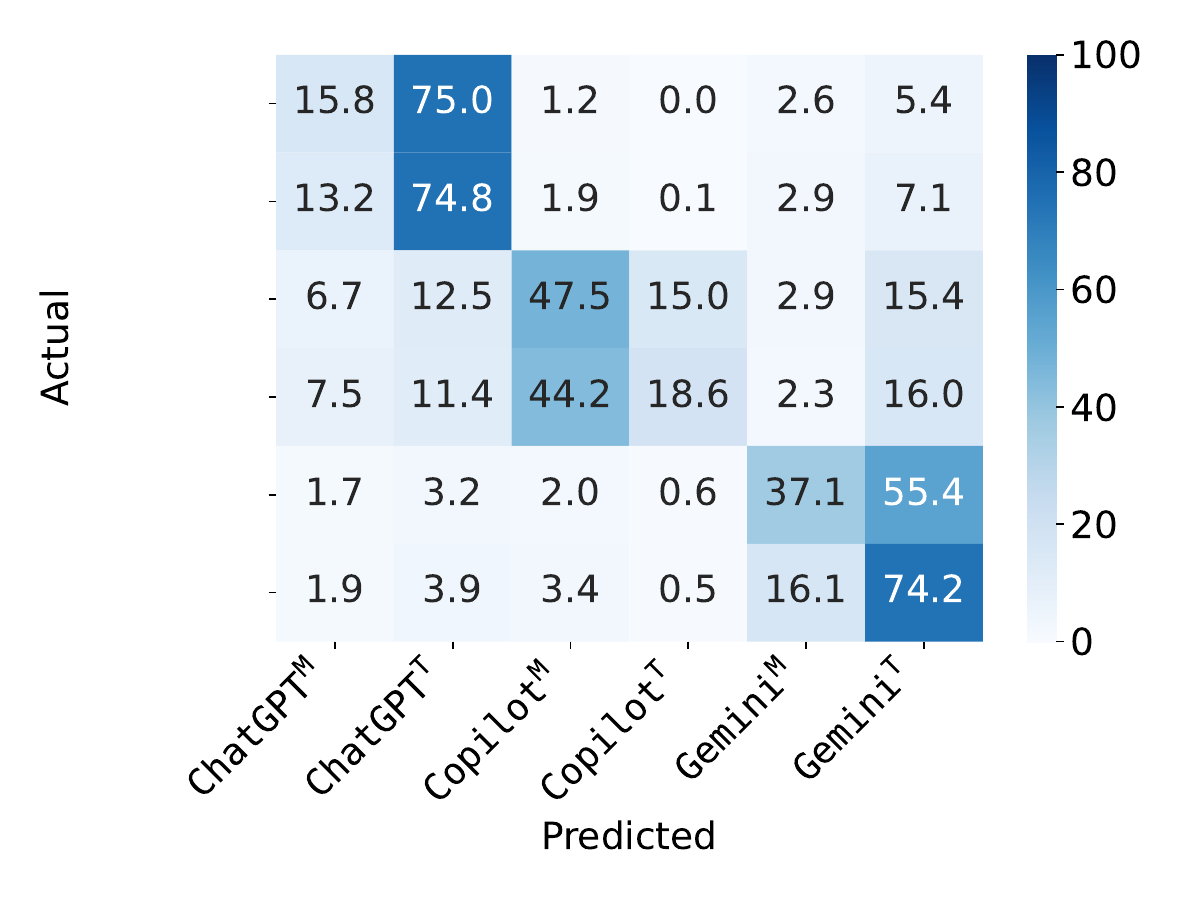}
    }

    \caption{Confusion matrices for GenAI app traffic classification tasks when \gls{sni} information is available at inference time versus when it is masked.}
    \label{fig:confusion_mat}
\end{figure}

\subsection{Traffic Classification of GenAI Apps and Content}
\label{subsec:classification}

In this section, we answer \emph{\gls{rq}3} by evaluating how effectively \gls{genai} apps and their generated content can be classified using traffic data. 
Building on the protocol-level analysis of Section~\ref{subsec:protocol} and prior findings in traffic classification~\cite{bayat2021deep,akbari2022traffic},
we also assess the contribution of the \gls{sni} extension to the \emph{classification of \gls{genai} app traffic}. 
To this end, we employ an \emph{occlusion analysis}~\cite{nascita2022machine}, a perturbation method which evaluates the impact on classification performance of masking the payload bytes corresponding to the \gls{sni} extension. 
We consider two different traffic-classification tasks: 
($i$) \emph{App Classification} ($3$ classes), which identifies the \gls{genai} app regardless of content type; 
and ($ii$) \emph{App\&Content Classification} ($6$ classes), which jointly distinguishes the \gls{genai} app and 
whether the generated content is textual or multimodal.

For both tasks, we adopt a lightweight \gls{onecnn} inspired by prior work on encrypted traffic classification~\cite{wang2017end,aceto2019mimetic,nascita2022machine}. 
The \gls{onecnn} takes as input the first $512$ raw payload bytes of the transport-layer payload of each biflow, encoded as 1D vectors. 
It applies two convolutional layers with $16$ and $32$ filters, respectively, and a kernel size of $25$. 
Each convolution layer is followed by ReLU activation and max pooling with a kernel size of $3$. 
The resulting feature maps are flattened and passed through a fully connected layer of $256$ units with dropout (rate = $0.2$) to mitigate overfitting. 
Finally, a dense layer with softmax activation outputs the class probabilities.

The occlusion analysis is performed at inference time: for each biflow, the bytes corresponding to the \gls{sni} extension are replaced with zeros (i.e.~a non-informative masking value). 
By comparing the performance with and without masking, we quantify the relative contribution of \gls{sni} information to classification accuracy. 

All experiments are repeated five times with different random seeds for the stratified train/test split. 
We report the mean and the standard deviation of the F1-score across the five repetitions.

App Classification achieves an F1-score of $(91.68 \pm 0.72)\%$, while the more challenging App\&Content Classification reaches $(49.90 \pm 1.40)\%$. 
When \gls{sni} values are masked through occlusion analysis, performance drops markedly to $(71.39 \pm 9.16)\%$ and $(38.89 \pm 4.81)\%$ F1-score for the two classification tasks, respectively. 
These results highlight the crucial role played by \gls{sni} information in both classification tasks.

To better understand these outcomes, we examine the average normalized confusion matrices reported in Fig.~\ref{fig:confusion_mat}. 
Figures~\ref{fig:cm_app} and~\ref{fig:cm_app_act} show the results for App Classification and App\&Content Classification, respectively, when \gls{sni} information is available at inference time. 
For the simpler App Classification task, inter-app confusion is very limited, with recall always above $88\%$ across all apps, consistently with the high F1-score obtained. 
Conversely, when performing App\&Content Classification, the model often confuses traffic generated by the same app across different content types, as highlighted by the squared-shaped error patterns. 
A plausible explanation is that multimodal responses frequently include textual descriptions in addition to images, which makes it difficult to discriminate between text and image generation. 
Nevertheless, even in this harder setting, the model still successfully distinguishes among apps.

When \gls{sni} information is masked, App Classification performance drops unevenly across apps (Fig.~\ref{fig:cm_app_masked}). 
In particular, \copilot recall drops to $52\%$ on average, with most biflows misclassified as \Gemini, while \chatgpt and \Gemini show only minor degradation. 
For App\&Content Classification under \gls{sni} masking (Fig.~\ref{fig:cm_app_act_masked}), confusion between content types within the same app becomes even more pronounced, especially for \chatgpt, whose multimodal traffic (\chatgptm) is mostly misclassified as text (\chatgptt). 
This confirms the reliance on \gls{sni} values that are strongly tied to specific content types, as also highlighted in Tab.~\ref{tab:sni} (e.g., \emph{files.oaiusercontent.com}, which exclusively accounts for $61\%$ of \chatgpt image-generation traffic). 
Moreover, inter-app confusion also increases: for instance, \copilot traffic is more frequently misclassified as \chatgpt or \Gemini. 
This indicates that, also when the classification task jointly considers both app and content type, \gls{sni} remains a substantial contributor to reliably distinguishing \gls{genai} traffic.

\begin{tcolorbox}[colback=gray!10,colframe=black,title=Answer to \gls{rq}3]
\gls{genai} apps and their generated content can be effectively classified from payload traffic, achieving $\approx\!92\%$ F1-score for app classification.
\gls{sni} plays a pivotal role in \gls{genai} traffic classification: when masked, performance drops markedly (up to $20$ percentage points) and misclassifications increase; inter-app confusion grows (e.g., lower \copilot recall with frequent misclassification as \Gemini), and within-app content-type errors become more pronounced (e.g., \chatgpt multimodal $\rightarrow$ text). 
Overall, \gls{sni} is a strong contributor, though not the sole discriminant, and its ongoing encryption poses practical challenges for traffic classification.
\end{tcolorbox}

\begin{figure*}[t]
    \centering

    \subfloat[\chatgpt (Down)]{
        \includegraphics[width=0.45\linewidth, trim={0 1.3cm 0 0}, clip]{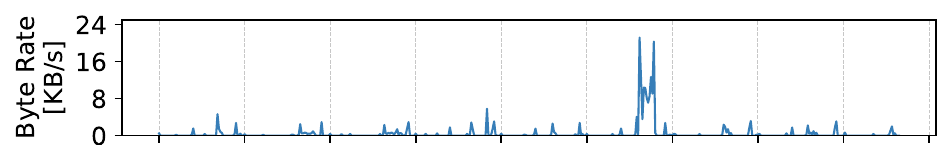}
    }\hfill
    \subfloat[\chatgpt (Up)]{
        \includegraphics[width=0.45\linewidth, trim={0 1.3cm 0 0}, clip]{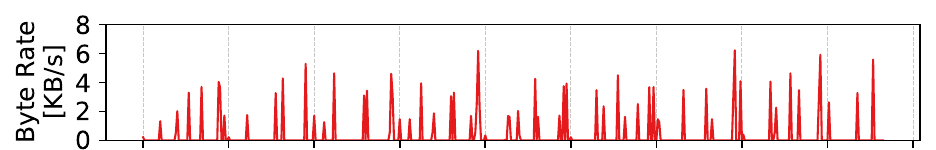}
    }
    
    \subfloat[\copilot (Down)]{
        \includegraphics[width=0.45\linewidth, trim={0 1.3cm 0 0}, clip]{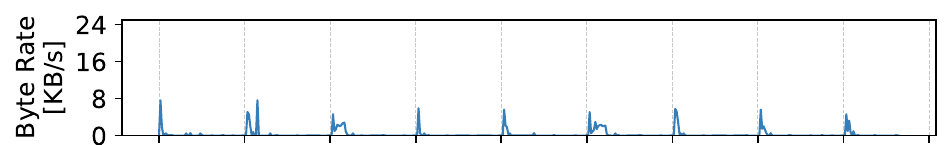}
    }\hfill
    \subfloat[\copilot (Up)]{
        \includegraphics[width=0.45\linewidth, trim={0 1.3cm 0 0}, clip]{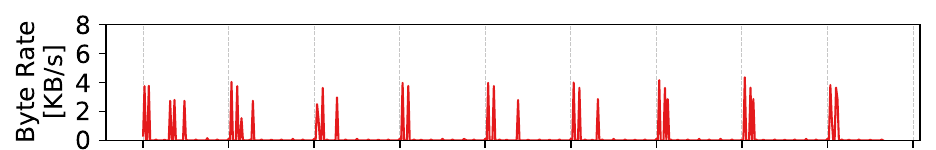}
    }
    
    \subfloat[\Gemini (Down)]{
        \includegraphics[width=0.45\linewidth, trim={0 1.3cm 0 0}, clip]{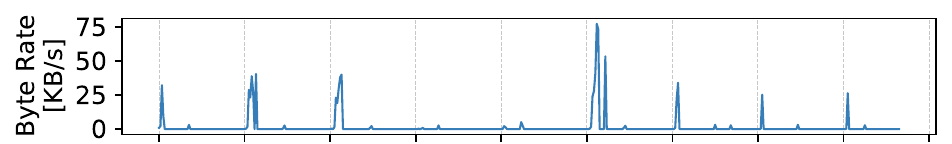}
    }\hfill
    \subfloat[\Gemini (Up)]{
        \includegraphics[width=0.45\linewidth, trim={0 1.3cm 0 0}, clip]{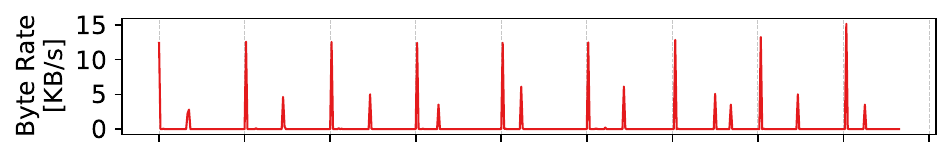}
    }
    
    \subfloat[\whatsapp (Down)]{
        \includegraphics[width=0.45\linewidth, trim={0 1.3cm 0 0}, clip]{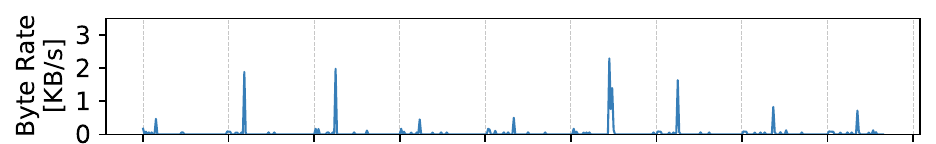}
    }\hfill
    \subfloat[\whatsapp (Up)]{
        \includegraphics[width=0.45\linewidth, trim={0 1.3cm 0 0}, clip]{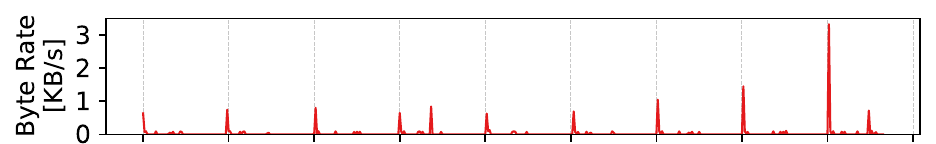}
    }
    
    \subfloat[\telegram (Down)]{
        \includegraphics[width=0.45\linewidth, trim={0 0cm 0 0}, clip]{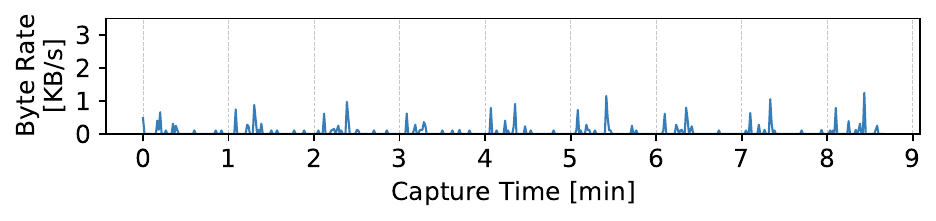}
    }\hfill
    \subfloat[\telegram (Up)]{
        \includegraphics[width=0.45\linewidth, trim={0 0cm 0 0}, clip]{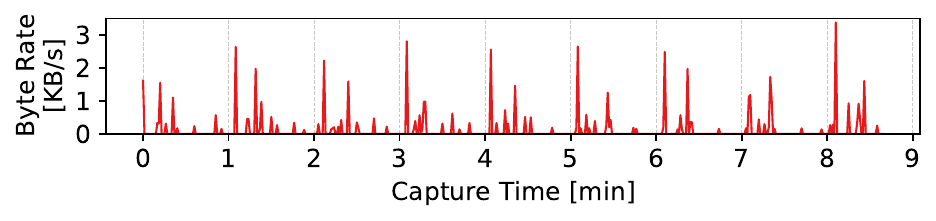}
    }
    
    \caption{
    Downstream (left column) and upstream (right column) traffic profiles of \gls{genai} chatbots (\chatgpt, \copilot, and \Gemini) and conventional messaging apps (\whatsapp and \telegram) when transmitting the prompts and corresponding \chatgpt  responses from the \emph{controlled} dataset. The traffic shown refers exclusively to prompts generating textual responses (i.e.~the first nine prompts).
    }
    \label{fig:volumetric_profiles}
\end{figure*}

\begin{figure*}[t]
    \centering

    \subfloat[\chatgpt (Down)]{
        \includegraphics[height=0.16\textwidth, trim={0cm 0cm 0 0}, clip]{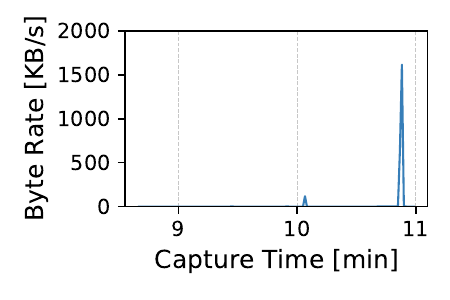}
    }
    \subfloat[\copilot (Down)]{
        \includegraphics[height=0.16\textwidth, trim={2cm 0cm 0 0}, clip]{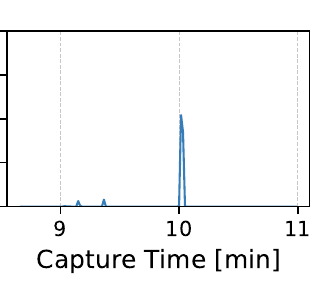}
    }
    \subfloat[\Gemini (Down)]{
        \includegraphics[height=0.16\textwidth, trim={2cm 0cm 0 0}, clip]{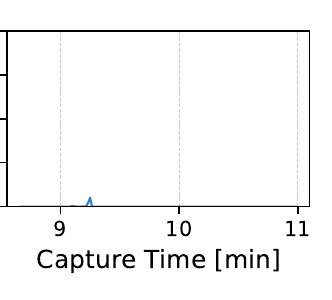}
    }
    \subfloat[\whatsapp (Down)]{
        \includegraphics[height=0.16\textwidth, trim={2cm 0cm 0 0}, clip]{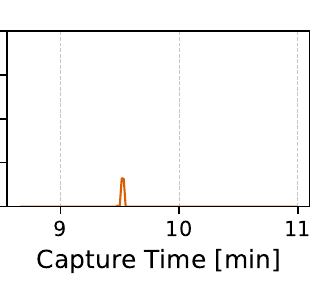}
    }
    \subfloat[\telegram (Down)]{
    \includegraphics[height=0.16\textwidth, trim={2cm 0cm 0 0}, clip]{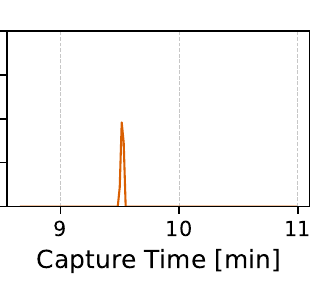}
    }
    
    \subfloat[\chatgpt (Up)]{
        \includegraphics[height=0.15\textwidth, trim={-0.5cm 0cm 0 0}, clip]{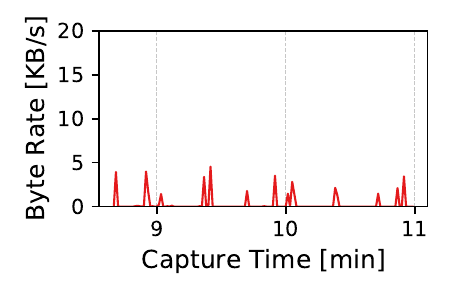}
    }
    \subfloat[\copilot (Up)]{
        \includegraphics[height=0.15\textwidth, trim={1.6cm 0cm 0 0}, clip]{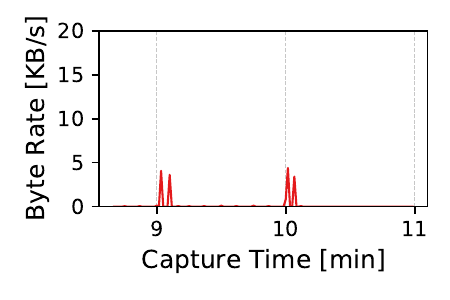}
    }
    \subfloat[\Gemini (Up)]{
        \includegraphics[height=0.15\textwidth, trim={1.6cm 0 0 0}, clip]{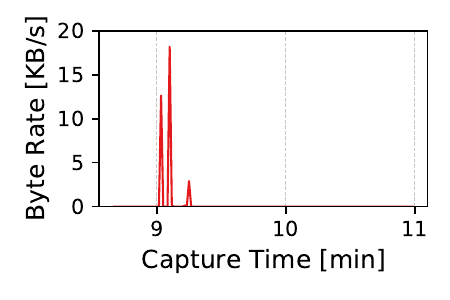}
    }
    \subfloat[\whatsapp (Up)]{
        \includegraphics[height=0.15\textwidth, trim={1.6cm 0cm 0 0}, clip]{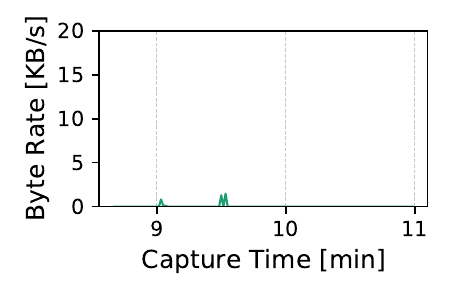}
    }
    \subfloat[\telegram (Up)]{
        \includegraphics[height=0.15\textwidth, trim={1.6cm 0cm 0 0}, clip]{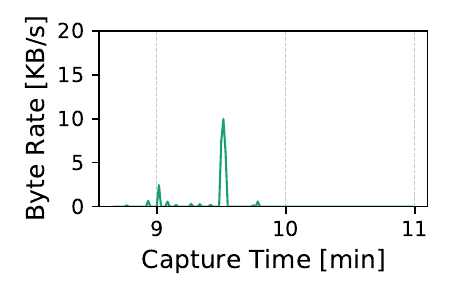}
    }
    
    \caption{
    Downstream (upper row) and upstream (lower row) traffic profiles of \gls{genai} chatbots (\chatgpt, \copilot, and \Gemini) and conventional messaging apps (\whatsapp and \telegram) when transmitting the prompts and corresponding \chatgpt responses from the \emph{controlled} dataset. 
    The traffic shown refers exclusively to the last (tenth) prompt, which generates an image-based response.
    }
    \label{fig:volumetric_profiles_immagine}
\end{figure*}

\subsection{Comparing the Traffic of GenAI Chatbots and Messaging Apps}
\label{subsec:controlled}

This section addresses \emph{\gls{rq}4} by exploiting the \emph{controlled} dataset to perform a fine-grained comparison of traffic profiles generated by \gls{genai} chatbots and conventional messaging apps (i.e.~\whatsapp and \telegram). 
Specifically, we consider the transmission of identical prompts and corresponding responses across all apps under the same capture conditions, enabling a direct comparison of their traffic behavior (see Section~\ref{subsec:controlled_prompts} for details). 
We distinguish between two scenarios based on the type of generated responses: ($i$) textual responses (corresponding to the first nine prompts) and ($ii$) multimodal responses involving image generation (tenth prompt). 

Figure~\ref{fig:volumetric_profiles} reports the downstream (left column) and upstream (right column) byte-rate profiles for \gls{genai} chatbots and conventional messaging apps when exchanging textual responses. 
When comparing \gls{genai} chatbots, \chatgpt and \copilot exhibit relatively modest downstream rates (below $8$~KB/s) across most prompts, whereas \Gemini shows pronounced downstream-rate peaks commonly up to $\approx\!40$~KB/s, despite generating a comparable amount of text for the same requests. 
A notable exception is Prompt~$6$ (request at minute~$5$), involving an online search%
\footnote{What cultural events are happening this month in Naples and Tokyo? Give me the response as plain text without images.}, 
where \chatgpt reaches $>\!20$~KB/s and \Gemini peaks at $\approx\!75$~KB/s, while \copilot maintains its usual rate. 
Response patterns also differ: \copilot and \Gemini transmit their responses almost immediately after the request (within the first $30$~s), whereas \chatgpt concentrates most of its downstream traffic in the latter half of the minute, probably delivering responses via incremental transmission of text chunks. 

Looking at the upstream profiles of \gls{genai} chatbots, \Gemini again exhibits the highest byte rates, with peaks up to $\approx\!10$~KB/s, compared to \chatgpt and \copilot, which never exceed $6$~KB/s. 
The prompt-transmission patterns also differ markedly: \chatgpt shows multiple upstream bursts throughout the entire minute, \copilot concentrates its peaks in the first half of each minute, while \Gemini displays an initial peak above $10$~KB/s followed by a second, smaller one. 
These observations reveal different transmission mechanisms and related traffic profiles for each \gls{genai} chatbot.

When comparing \gls{genai} chatbots with conventional messaging apps, we observe markedly different traffic profiles in both downstream and upstream directions.
All reported results correspond to interactions based on \chatgpt responses, which we adopted as reference since comparable behavior was observed when replicating responses of the other chatbots.
Messaging apps generate significantly less traffic: downstream rates remain below $2$~KB/s for \whatsapp and below $1$~KB/s for \telegram, while upstream rates are always $<\!3$~KB/s for both. 
In both directions, the traffic profiles are consistent with the amount of text exchanged. 
On the other hand, \whatsapp exhibits a highly peaked pattern, with bursts occurring only at transmission times, whereas \telegram shows a persistent low-rate background traffic throughout the entire time interval. 

Multimodal content generation, depicted in Fig.~\ref{fig:volumetric_profiles_immagine}, also exhibits distinctive traffic profiles both across \gls{genai} chatbots and when compared to messaging apps. 
In the downstream direction (up row), all three \gls{genai} chatbots show an initial small peak followed by a short burst of high-rate traffic, significantly exceeding the rates observed for text generation. 
The first peak likely corresponds to the transmission of the textual caption accompanying the image, while the subsequent high-rate burst might be associated with the delivery of the image itself. 
Conversely, messaging apps display a single downstream peak corresponding to the download of the transmitted image, with \whatsapp showing a slightly higher rate than \telegram, as also observed in the text-generation scenario. 
Upstream traffic profiles reveal distinct behaviors that are peculiar to each \gls{genai} and messaging app. 
They closely mirror those observed in the textual generation, since the prompt is purely textual, also in the image-generation scenario.
Overall, \gls{genai} chatbots impose a much higher load on the network, particularly in the upstream direction. 
While high downstream rates are expected for apps delivering large content, as in the case of image transmission (regardless of whether the content is AI-generated), the sustained and often substantial upstream rates represent a \emph{novel stress factor for mobile networks introduced by \gls{genai} workloads}. 
This finding reinforces the urgent challenge highlighted in recent industry reports~\cite{ericsson2025}.

\begin{tcolorbox}
[colback=gray!10,colframe=black,title=Answer to \gls{rq}4]
Using the \emph{controlled} dataset with identical content across apps, we observe marked differences between \gls{genai} chatbots and conventional messaging apps in both downstream and upstream traffic profiles. 
\gls{genai} chatbots not only generate higher downstream rates but also sustain elevated upstream activity, even for purely textual prompts.
\chatgpt, \copilot, and \Gemini exhibit distinct temporal profiles and transmission strategies; whereas \whatsapp shows short, isolated bursts and \telegram maintains a persistent low-rate background traffic. 
These differences hold for both text-only and image-generation requests, suggesting that the nature of \gls{genai} processing, rather than the prompt format, drives the observed traffic patterns.
\end{tcolorbox}

\section{Related Work}
\label{sec:bg}

Building on their success, the content generated by \gls{genai} chatbots has been characterized and evaluated across a wide range of application domains (e.g., financial decision-making~\cite{roychowdhury2024journey}, code generation~\cite{liu2023advances}, and scientific literature search~\cite{gwon2024use}) and from multiple perspectives (e.g., technological comparison~\cite{rahman2025comparative}, ethical and quality concerns~\cite{kim2023you}, and broader social risks~\cite{yang2024social}).

\begin{draft}
In contrast, their network-level behavior remains comparatively less explored. 
\citet{lyu2024measuring} investigate the network characteristics of \gls{genai} apps. 
However, unlike the present work, their primary goal is to infer usage patterns and activity trends across various \gls{genai} platforms in a campus network. 
Their methodology relies on \gls{sni}-based measurements to associate network flows with specific \gls{genai} apps and correlate the resulting traffic patterns with temporal campus activities.
More recently, \citet{cheng2025hello} present a measurement study of human-to-\gls{genai} calling applications.
While their observations on transport protocols (e.g., the adoption of QUIC by \Gemini) align with our findings, their work specifically targets real-time voice-based conversational systems, focusing on audio latency and interaction dynamics.
Our work differs from these latter studies in both scope and objectives. Rather than analyzing usage trends or voice-calling systems, we conduct an in-depth traffic characterization of \gls{genai} chatbot apps. Specifically, we focus on the distinct dynamics of textual and multimodal generation, analyzing protocol footprints, packet-sequence dynamics, and traffic visibility mechanisms,  with particular emphasis on implications for network management.
\end{draft}

On the other hand, several studies conduct network traffic characterization and modeling across different application domains and operational scenarios. These include video streaming services~\cite{gharakheili2019itelescope, madanapalli2021modeling, wang2024characterizing}, virtual reality platforms~\cite{lyu2023metavradar}, online gaming~\cite{lyu2024network,carracosa2022cloud}, and communication and collaboration apps~\cite{guarino2021characterizing, michel2022enabling}. 
Other prior work~\cite{aceto2021characterization, sasidharan2021prodroid, guarino2021characterizing} models the network behavior of various kinds of apps using Markovian models in addition to volumetric and statistical properties.
\begin{draft}
However, existing traffic characterization and modeling studies do not explicitly address the emerging class of \gls{genai} chatbot applications, whose inference-driven and cloud-centric workloads introduce distinct traffic patterns and network stress factors.
\end{draft}

\section{Conclusions and Future Perspectives}
\label{sec:end}
Understanding the characteristics and distinct patterns of traffic is essential for various networking tasks, including network monitoring, management, planning, provisioning, as well as to ensure network security and robustness.
This work provided a characterization and modeling of the network traffic generated by \gls{genai} chatbots accessed via their mobile apps, focusing specifically on \chatgpt, \copilot, and \Gemini. 
We demonstrated that the traffic generated by these \gls{genai} chatbots constitutes a new category compared to existing mobile traffic, underlining the need to understand how such workloads may reshape network dynamics and influence operational practices. 
To this aim, we collected more than $60$ hours of human-generated traffic using both generic (i.e.~unconstrained) and controlled prompts, and we publicly released the resulting $\mathtt{MIRAGE\text{-}GenAI\text{-}2025}$ datasets to foster reproducibility and further research.%
\footref{foot:dataset}

The outcome of our analysis revealed distinct network behaviors among \gls{genai} chatbots, influenced by both the app and the type of generated content, and it directly answered a set of practical research questions. 
(\emph{\gls{rq}1}) \copilot and \Gemini exhibited a higher network load during multimodal generation, whereas the likely incremental transmission of \chatgpt in groups of tokens led to increased traffic even for purely textual responses.
Flow-level characterization and modeling highlighted differences in payload size and inter-arrival time distributions, with \chatgpt showing significant download activity and large payload sizes, while \copilot and \Gemini exhibited more diverse dynamics dependent on the generated content.
(\emph{\gls{rq}2}) At the protocol level, we observed differences in transport protocol adoption (e.g., \Gemini's extensive usage of QUIC), \gls{tls} version (e.g., \chatgpt exclusively relying on \gls{tls}\;1.3), and \gls{sni} values. 
(\emph{\gls{rq}3}) The latter proved highly discriminative for distinguishing both apps and content types, with \gls{sni} masking leading to a marked performance drop and increased misclassifications, as revealed by our occlusion analysis. 
(\emph{\gls{rq}4}) Finally, by comparing \gls{genai} chatbots and conventional messaging apps when carrying the same content, we uncovered substantial differences in both downstream and upstream traffic profiles: \gls{genai} chatbots combined heavy downstream usage with high upstream activity, even for text-only prompts, introducing a novel stress factor for mobile networks.

These findings underscore the importance of characterizing \gls{genai} traffic as a necessary step to anticipate its impact on network usage. 
Such knowledge is essential to guide network monitoring, planning, and resource management, ultimately helping operators prepare their infrastructures for the widespread adoption of \gls{genai}-based services.

\begin{draft}
Future research will leverage this knowledge for in-depth classification and fine-grained prediction of \gls{genai} chatbot traffic with advanced machine learning and deep learning approaches.
Furthermore, we plan to expand our dataset by capturing traffic from additional chatbots (e.g., Claude and DeepSeek) and covering other content modalities (e.g., video and audio), while also performing comparative analyses (e.g., with common search engines) to deepen our understanding of these emerging applications.
In addition, we intend to develop fully automated frameworks for prompt injection and response collection to enable larger-scale and repeatable measurement campaigns. 
Finally, we aim to extend our measurement infrastructure to support multiple vantage points, enabling traffic collection from geographically distributed locations to investigate the impact of device location on observed traffic patterns.
\end{draft}

\section*{Acknowledgements}
This research has been partially carried out within the ``xInternet'' Project supported by the MUR PRIN 2022 program (D.D.104---02/02/2022) funded by the NextGenerationEU. 
This manuscript reflects only the authors' views and opinions, and the Ministry cannot be considered responsible for them.

\balance
\bibliographystyle{unsrtnat}
\begingroup
\footnotesize
\bibliography{bibliography}
\endgroup

\appendix
\section{Ethical Considerations}
\label{appendix:ethics}
\noindent\textbf{Traffic Data Collection.}
This study involves the collection and analysis of network traffic generated by \gls{genai} chatbot apps. Traffic collection was conducted in a controlled environment using experimental smartphones, rather than the personal devices of users connected to the Internet, as detailed in Sec.~\ref{subsec:capture_setup}. The chatbots were accessed via purpose-created accounts specifically set up for the data collection campaign to avoid any use of real user credentials or personal data.
All experimenters participating in the capture sessions were fully informed about the nature and purpose of the experiments beforehand and provided voluntary consent. They agreed that the collected network traffic data could be used and shared for research purposes. No personally identifiable information or sensitive content was collected, and all data has been handled following principles of minimal risk and respect for privacy.

\vspace{5pt}
\noindent\textbf{Use of Generative AI Tools.}
Regarding the writing of the manuscript, Generative AI tools and technologies were employed exclusively for grammatical support and minor text polishing.

\end{document}